\documentclass[a4paper,fleqn]{cas-sc}

\usepackage[authoryear,longnamesfirst]{natbib}

\usepackage{amssymb}
\usepackage{amsmath}
\usepackage{subcaption}
\usepackage{url}
\usepackage{xurl}
\usepackage{array}
\usepackage{longtable}
\usepackage{float}
\usepackage{xcolor}

\usepackage{placeins}
\let\origsection\section
\renewcommand{\section}{\FloatBarrier\origsection}

\begin{document}
\let\WriteBookmarks\relax
\def\floatpagepagefraction{1}
\def\textpagefraction{.001}

\shorttitle{Bayesian deep learning for 3D copper targeting at Kogodai}
\shortauthors{M.~Veshchezerova et~al.}

\title[mode = title]{Bayesian deep learning integration of geophysical and drilling data for 3D prediction of copper mineralization and drill targeting: a case study from the Kogodai prospect, Rudny Altai}

\author[1]{Margarita Veshchezerova}
\credit{Conceptualization, Methodology, Software, Formal analysis, Writing -- original draft}

\author[1]{Egor Barashov}[orcid=0000-0002-4335-1615]
\cormark[1]
\ead{eb@terraquantum.swiss}
\credit{Methodology, Software, Investigation, Visualization, Writing -- original draft}

\author[1]{Evgenii Gusev}
\credit{Software, Validation, Data curation}

\author[1]{Michael R. Perelshtein}
\credit{Supervision, Project administration, Writing -- review \& editing}

\affiliation[1]{organization={Terra Quantum AG},
            addressline={Kornhausstrasse 25},
            postcode={9000},
            city={St.~Gallen},
            country={Switzerland}}

\author[2]{Arlan Kasymzhan}
\credit{Resources, Investigation, Data curation}

\author[2]{Bolat M. Kabaziev}
\credit{Resources, Supervision}

\author[2]{Nurlan Y. Askarov}
\credit{Resources, Data curation}

\affiliation[2]{organization={Kogodai Joint Venture LLP},
            addressline={Gorky Street 46},
            postcode={070004},
            city={Ust-Kamenogorsk},
            country={Kazakhstan}}

\cortext[1]{Corresponding author.}

\begin{abstract}
Exploration drill targeting in structurally complex terranes is hindered by sparse sampling, heterogeneous datasets, and the ambiguity of geophysical inversions. Here, we present an uncertainty-aware 3D workflow for the acceleration of time-to-discovery in brownfield explorations and apply it to the Kogodai prospect in the Rudny Altai metallogenic province. We jointly analyse existing drilling and geophysical data in a comprehensive approach, revealing hidden patterns in already available data. Drillholes and trenches were desurveyed to a common 3D reference frame, and assays were composited to a consistent spatial support to facilitate joint modelling with geophysical inputs.
We develop Bayesian deep-learning models to predict 3D fields of Cu grade together with chargeability and apparent resistivity while quantifying epistemic uncertainty via Monte Carlo sampling. The original contribution of this work is to treat the problem not as pointwise regression between co-located observations, but as joint learning of spatially continuous 3D fields from sparse, heterogeneous exploration evidence. The resulting 3D predictions delineate a principal mineralized trend and several localized candidate zones that coincide with elevated induced polarization (IP) responses, while uncertainty mapping highlights where predictions are robust versus where additional drilling would be most informative. The continuous Cu-grade field can also be thresholded to produce binary prospectivity maps, allowing the sensitivity of target delineation to the chosen cutoff to be evaluated. The outputs are intended for qualitative interpretation and risk-aware drill targeting rather than resource estimation, and we discuss key limitations arising from incomplete provenance metadata for geophysical products and heterogeneity of historical sampling.
\end{abstract}



\maketitle

\section{Introduction}
\label{sec1}

\begin{figure}[!htbp]
    \centering
    \includegraphics[width=0.55\textwidth]{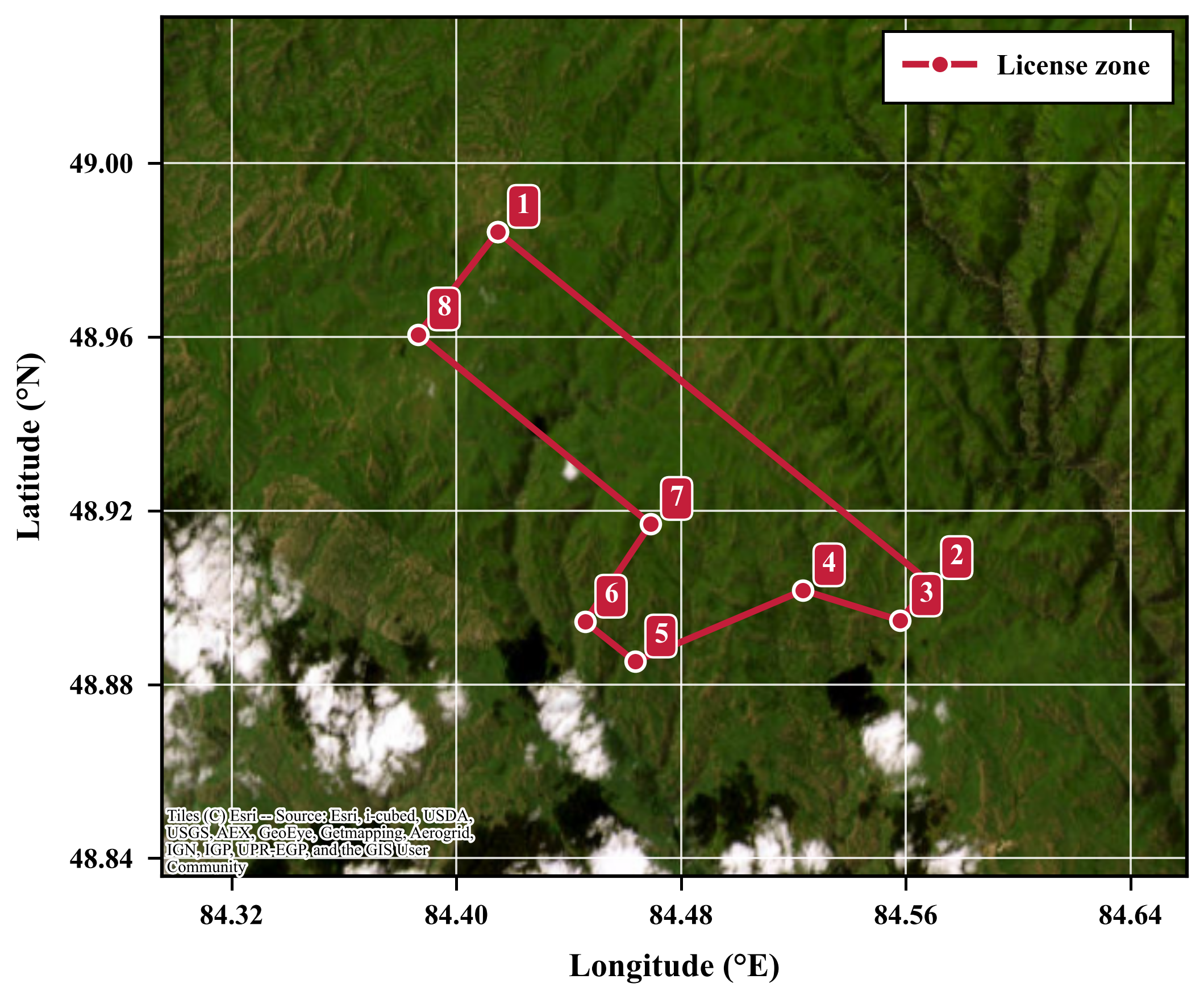}
    \caption{Satellite view of the Kogodai licence area, eastern Kazakhstan. The licence perimeter contains the Kogodai Cu-rich VMS / copper massive-sulfide deposit and the Lotoshnoye occurrence, both hosted by high-grade metamorphic rocks of the Kurchum block within the Irtysh shear zone.}
    \label{fig:licence_perimeter}
\end{figure}
\begin{figure}[!htbp]
    \centering
    \includegraphics[width=\textwidth]{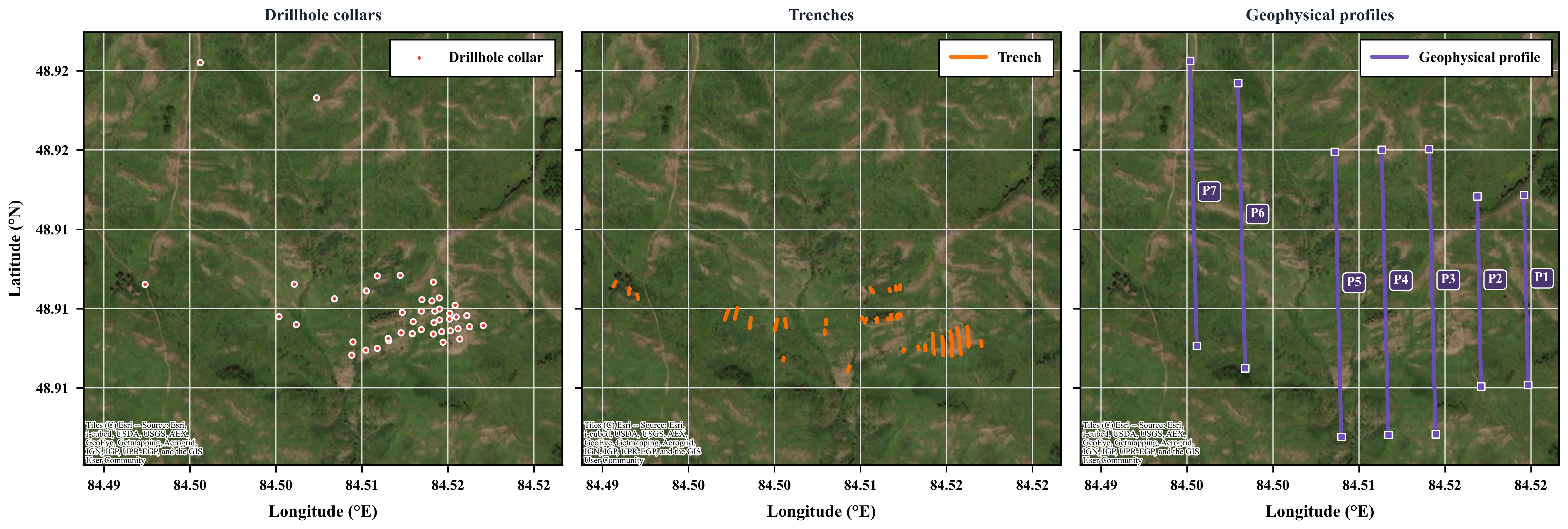}
    \caption{Collected data locations: drillholes, trenches and geophysical profiles from left to right.}
    \label{fig:collected_data_locations}
\end{figure}

\textit{Exploration drilling} is the principal source of information on subsurface mineralization, but it is also one of the most expensive processes in mineral exploration. In \textit{brownfield settings}, where a junior exploration company seeks to extend known resources and identify nearby satellite bodies, the practical problem is rarely a lack of data alone. More commonly, the challenge is to integrate sparse drilling, historical maps, trench information and indirect geophysical measurements into a coherent \textit{3D interpretation} that can guide the next drilling campaign. The present study addresses such a problem at the Kogodai copper prospect (Figure~\ref{fig:licence_perimeter}), where the available dataset includes historical geological information, 45 drillholes, trenches, and \textit{induced-polarization} (IP) and \textit{resistivity} profiles (Figure~\ref{fig:collected_data_locations}). The objective is not to produce a mineral resource estimate, but to support exploration-drilling planning by identifying zones where copper mineralization is probable and where additional drilling would most reduce uncertainty.

This abundance of data and the need for data integration motivates the use of \textit{artificial intelligence} (AI) in brownfield targeting. Neural networks can represent nonlinear relationships between heterogeneous inputs and mineralization that are difficult to express as explicit geological rules, while in practical exploration they can act as an \textit{independent second opinion}: validating known trends, highlighting overlooked anomalies, ranking targets and reducing the risk that useful patterns in legacy datasets are missed. Industry case studies, such as VRIFY's DORA system on Canterra Minerals' Buchans Project, illustrate this demand for AI-assisted target validation and ranking.\footnote{See the VRIFY Canterra Minerals case study: \url{https://vrify.com/case-studies/canterra-minerals-enhanced-historic-exploration-work-with-ai}.} Such examples should not replace peer-reviewed evidence, but they reflect the practical demand for reproducible and data-driven target generation.

\subsection{Prospectivity modelling as spatial evidence integration}
This exploration drilling planning is closely related to \textbf{mineral prospectivity mapping} (MPM), which integrates \textit{evidential features} derived from multi-source geoscientific spatial datasets in order to delineate and rank prospective areas for undiscovered mineral deposits of a given type \citep{bonhamcarter1994gis,carranza2010predictive,fu20213d}. Classical MPM workflows may be \textit{knowledge-driven} (based on expert geological criteria) or \textit{data-driven}, when the spatial association between known mineral occurrences and evidential layers is learned statistically or by machine learning \citep{carranza2008geochemical,carranzaLaborte2015random,fu20213d}. In both cases, MPM typically operates on a \textit{cell-by-cell} or \textit{block-by-block} basis: each spatial unit is represented by a vector of evidential features, such as proximity to structures, geophysical anomalies, geochemical signatures, lithological contacts or alteration proxies, and the model estimates a prospectivity value for that unit. Spatial continuity may enter indirectly through the construction of evidential layers, for example by interpolating geochemical samples or by calculating distances and densities of structural features. The prediction itself, however, is generally formulated as a supervised or rule-based association between cell feature vectors and known mineralized or non-mineralized locations.

Traditionally, MPM has been applied at regional to belt scales, where the goal is to reduce a large search space and identify areas for follow-up exploration. The translation of MPM to \textit{target, camp or deposit scales} is less straightforward because not all mineral-system components remain resolvable in local datasets. In the Rajapalot Au--Co project, \citet{chudasama2022target1,chudasama2022target2} demonstrate this target-scale adaptation by combining fuzzy, neuro-fuzzy, self-organizing-map and neural-network approaches to quantify model uncertainty, identify deposit-related patterns and prioritize local drill targets.

Prospectivity modelling has also been extended into 3D: \citet{nielsen2019tampia} used geological, geophysical, geochemical and rock-property data to build a \textit{3D mineral potential model} that informed resource-domain interpretation, while \citet{fu20213d} trained machine-learning classifiers on 3D geological and geophysical inversion products at the Zhuxi tungsten deposit. These studies show that 3D prospectivity modelling can integrate geological and geophysical data, but it remains primarily a probability or classification framework rather than a direct grade-prediction framework.

\subsection{Grade estimation and neural-network resource models}
The related problem of estimating grades in unsampled locations is classically addressed by \textbf{mineral resource estimation} (MRE). Since \citet{matheron1971theory}, \textit{kriging} has become a standard tool for estimating grades from drilling and assay data. In modern resource workflows, grade interpolation is typically constrained by geological domains, structural interpretations and mineralized envelopes. This makes MRE different from MPM in both purpose and scope. MPM asks where mineralization is geologically favourable or likely to occur; MRE estimates grade and endowment within a modelled resource domain. The distinction is not absolute, however. \citet{nielsen2019tampia} show that mineral potential models may help inform or constrain resource domains, while resource models provide grade estimates that can be compared against prospectivity predictions.

Deep-learning resource models have recently been explored at data-rich mine sites: at Jundee, \citet{first2023introducing} used \textit{convolutional neural networks} to model narrow high-grade gold veins and nonlinear multielement patterns, improving \textit{precision} or \textit{recall} relative to kriging. Such workflows rely on dense underground drilling, grade-control and rock-chip data, whereas early- to mid-stage exploration datasets are sparse, uneven and require a more uncertainty-aware formulation.

\subsection{Drill planning as sequential decision-making}
A third related line of work concerns optimal drilling and sequential data acquisition. \citet{mern2022intelligent} formulate mineral exploration as a \textbf{sequential decision-making problem under uncertainty}. Their \textit{Intelligent Prospector} framework represents the evolving state of geological knowledge as a belief and uses a \textit{partially observable Markov decision process} to select measurements that improve predictions and decisions. \citet{scheidt2026optimizing} work on \textit{efficacy of information} (EOI) frame drillhole ranking in terms of expected uncertainty reduction in relevant economic quantities such as grade, tonnage or volume. In this approach, a set of plausible prior geological or grade models is generated, candidate drillholes are simulated through these prior models, and the expected information gain from each drillhole is evaluated. These approaches are directly relevant to exploration planning because they do not merely predict where mineralization may occur; they also ask which new measurement is expected to be most valuable for reducing geological and economic uncertainty.

\subsection{Contribution of this study}
In this study, we develop an \textbf{uncertainty-aware 3D neural-network model} for the Kogodai copper prospect, a \textit{VMS-related copper occurrence} in the Kurchum block of eastern Kazakhstan. Hosted in highly metamorphosed rocks of the Irtysh shear-zone framework, Kogodai has mineralization localized near amphibolite--gneiss/terrigenous-rock contacts and controlled by structural-tectonic factors \citep{bekbotayeva2023features}.

Our approach combines elements of MPM and MRE while remaining distinct from both. As in MRE, the model uses known assay locations to learn a spatially continuous prediction of copper grade in unsampled blocks. As in MPM, it also uses additional geophysical features, especially IP chargeability and resistivity, whose spatial association with mineralization can support prediction away from direct sampling. Unlike conventional MPM, the model does not estimate a probability of mineralization from predefined evidential layers, but rather predicts continuous 3D fields of copper grade and geophysical properties. Unlike conventional MRE, it is not intended to produce a resource-grade block model or grade-tonnage estimate, but to support exploration targeting and drill planning. Nevertheless, the predicted grade field can be interpreted as a potential model if a cutoff grade is selected and blocks with predicted grades above this cutoff are treated as prospective.

Unlike target-scale MPM studies that address model or systemic uncertainty in prospectivity values \citep{chudasama2022target1}, our Bayesian formulation estimates \textbf{epistemic uncertainty} from limited and uneven observations by sampling neural-network weights \citep{he2025survey}, producing \textit{predictive means} and \textit{uncertainty estimates} that separate robust targets from weakly constrained anomalies.

The resulting workflow provides a practical \textbf{decision-support tool} for brownfield exploration at Kogodai. It helped validate known mineralization trends and identify candidate zones that may have been overlooked in earlier interpretations. Depending on the company's strategic objective, the outputs can be used in two complementary ways: to accelerate time to discovery by prioritizing high-probability copper targets, or to reduce exploration risk by selecting drillholes expected to most improve confidence in the subsurface interpretation. The study therefore contributes to the emerging interface between 3D prospectivity modelling, machine-learning-assisted grade prediction and uncertainty-aware drill planning.

\section{Geological background}
\label{sec2}

\subsection{Regional setting}

\begin{figure}[!htbp]
    \centering
    \includegraphics[width=\textwidth]{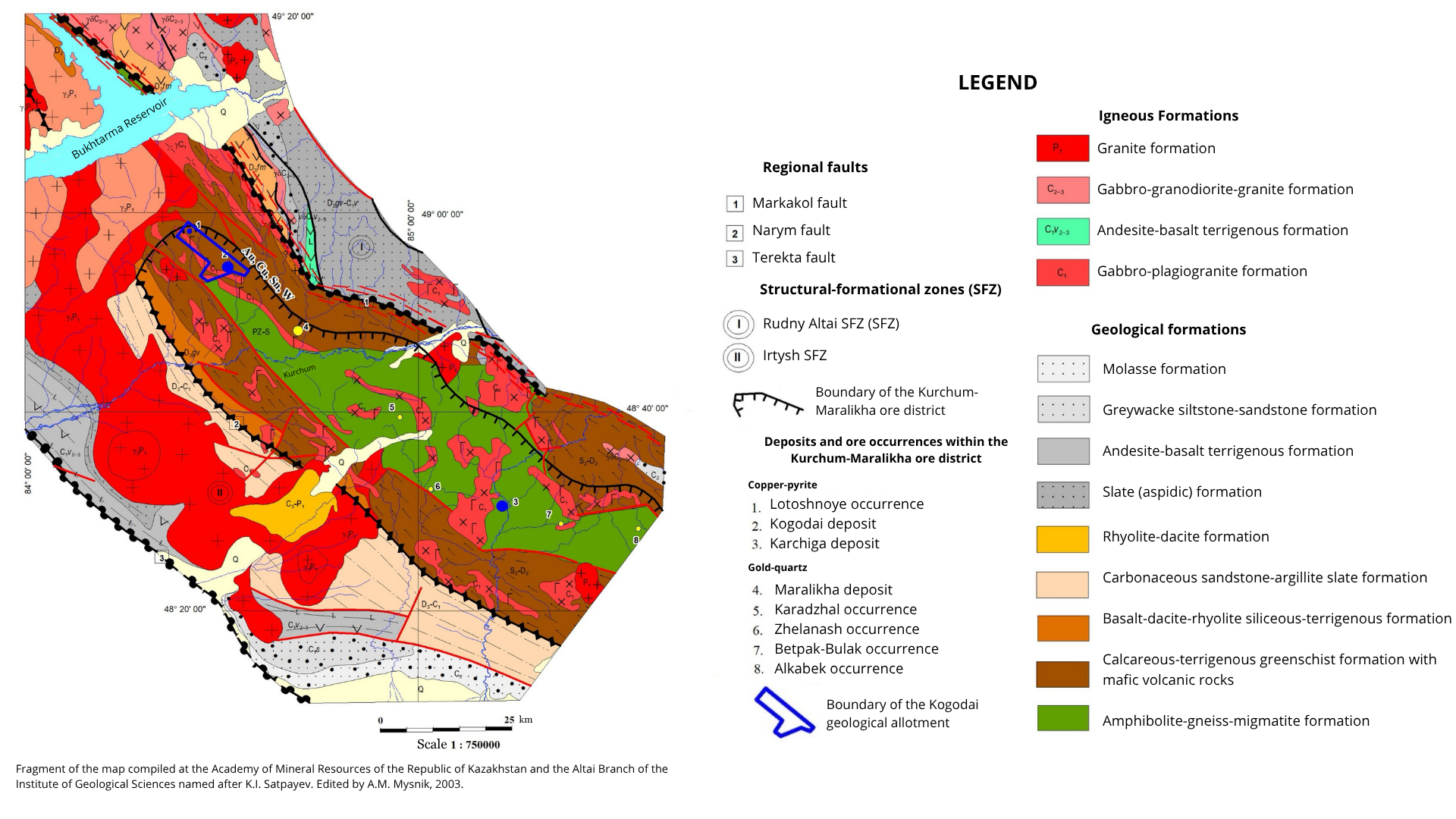}
    \caption{Regional geological map of the Kogodai area, showing the broader lithological and structural context of the prospect within the Rudny Altai metallogenic province.}
    \label{fig:regional_geological_map}
\end{figure}

The Kogodai prospect is located in the Rudny Altai metallogenic province (Figure~\ref{fig:regional_geological_map}), a region well known for volcanogenic massive sulfide (VMS) and related Cu--Zn--Pb mineralization formed during Paleozoic tectono-magmatic evolution of the Altai--Mongolian fold belt \citep{kozlov2015formation,vikentyev2024rudny}. Recent metallogenic syntheses emphasize the role of Devonian bimodal basalt--rhyolite volcanism developed in an extensional to transtensional regime, including pull-apart basins, and highlight systematic along-strike metallogenic zoning within the belt \citep{vikentyev2024rudny}. The metallogenic development of the province has been linked to arc/back-arc magmatism, elevated heat and fluid fluxes, and the formation of pyrite-bearing hydrothermal--sedimentary horizons and VMS/Cu--VMS deposits, with several stages of tectonomagmatic evolution recognized through the Silurian--Carboniferous interval \citep{kozlov2015formation}. Comparative analyses of massive-sulfide provinces indicate that, relative to the Southern Urals, Rudny Altai is characterized by more felsic (rhyolitoid) island-arc magmatism and a predominance of polymetallic massive-sulfide mineralization, with distinct vertical and lateral distribution patterns \citep{seravkin2019southern}.

Isotopic constraints further indicate that the Rudny Altai VMS systems represent a large and coherent province with regionally consistent metal sources; high-precision Pb-isotope data from numerous deposits support a largely homogeneous, predominantly lithospheric mantle--related lead source rather than a juvenile asthenospheric signature \citep{chernyshev2023sources}. At the scale of the Greater Altai region, regional syntheses propose structural (fault-controlled), lithostratigraphic, and magmatic criteria that can support mineral exploration targeting \citep{d2022geological}.

\subsection{Local geological framework}

\begin{figure}[!htbp]
    \centering
    \includegraphics[width=\textwidth]{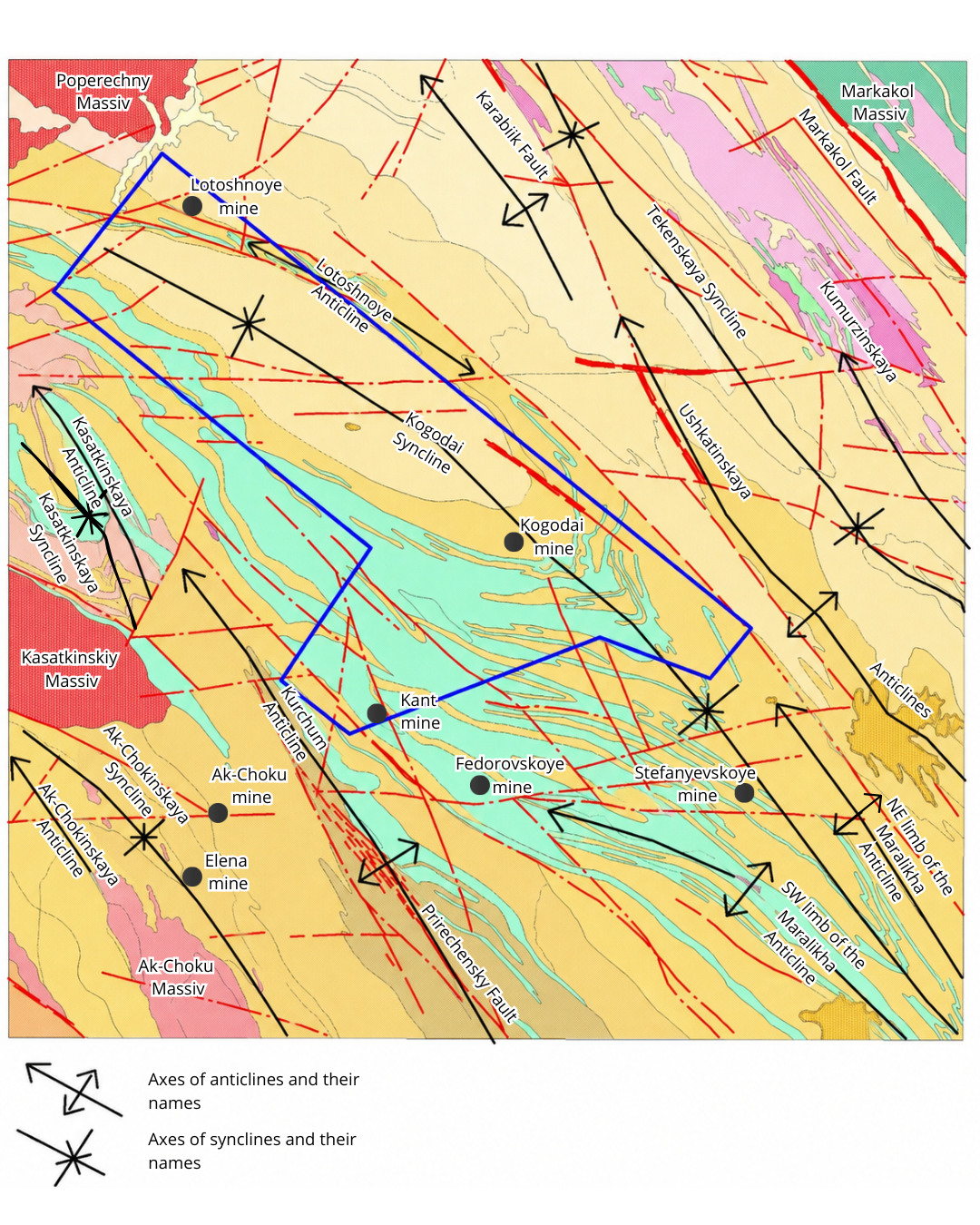}
    \caption{Tectonic map of the Kogodai--Lotoshnoye area showing its position within the Axial Subzone of the Irtysh structural-formational zone and the distribution of second-order folds (Kogodai syncline and Lotoshnoye anticline).}
    \label{fig:kogoday_lotoshnoye_tectonic_map}
\end{figure}

The Kogodai prospect is situated within the northwestern part of the Kurchum block of high-grade metamorphic rocks, within the broader Irtysh shear zone. The prospect is associated with the centroclinal closure of the Kogodai syncline (Figure~\ref{fig:kogoday_lotoshnoye_tectonic_map}), which trends NW (approximately 300--310$^{\circ}$) and plunges gently (15--20$^{\circ}$) to the northwest. Minor longitudinal and transverse faults are reported, but large offsets are not dominant at the scale of the prospect.

The Irtysh shear zone represents a major transregional structure within the Central Asian Orogenic Belt and records multiple episodes of deformation and reactivation, including late Paleozoic collision-related evolution and later Mesozoic--Cenozoic reactivation \citep{glorie2012tectonic}. This tectonic framework is important for understanding structural pathways and the juxtaposition of lithological packages in the Kurchum block and, consequently, for interpreting mineralization controls in high-grade metamorphic host rocks.

The local stratigraphy is dominated by high-grade metamorphic lithologies, including feldspar--quartz--mica schists and gneisses, with frequent amphibolite horizons. Amphibolites occur as massive and banded bodies and are accompanied by subordinate quartz--feldspar--mica and amphibole schists. Small concordant amphibolite bodies and granitoid intrusions are present, including granodiorite-porphyry dikes that locally crosscut fold-related fabrics. In the Kogodai area, mineralized zones are commonly associated with contacts between amphibolite bodies and gneissic/terrigenous metamorphic rocks, and the spatial juxtaposition of lithologies of different ages and compositions reflects intense metamorphic overprinting and structural reworking \citep{bekbotayeva2023features}.

\subsection{Mineralization style and controls}

Copper mineralization at Kogodai is classified as a copper--pyrite (Cu--VMS-related) style and is broadly comparable to other deposits within the Rudny Altai province. Across the prospect, mineralized zones are typically lens- to ribbon-shaped and broadly conformable with the host-rock fabric; at a regional scale, similar copper--pyrite occurrences are widely reported to be preferentially localized along amphibolite--gneiss contacts within high-grade terrigenous metamorphic sequences. This style is consistent with other Cu-bearing massive sulfide occurrences hosted by high-grade metamorphic complexes within the Kurchum block, where mineralization is spatially linked to amphibolite horizons and associated terrigenous rocks \citep{lobanov2012karchiga}.

Mineralization is commonly developed within zones of schistosity and deformation along lithological contacts, and is characterized by pyrite, chalcopyrite, pyrrhotite and sphalerite. Alteration associated with mineralized zones may include chloritization and epidote--actinolite assemblages, locally accompanied by talc, consistent with hydrothermal metasomatism in contact- and shear-controlled settings \citep{bekbotayeva2023features}. Structural controls include both folding-related features (e.g., hinges and closures) and brittle structures; local surface faults are interpreted as potential pathways for mineralizing fluids \citep{bekbotayeva2023features}.

At the prospect scale, five zones of hydrothermally altered rocks with visible copper mineralization have been reported. The largest zone occurs in the southern part of the area and extends for more than 500~m along strike. The central part of the prospect is relatively better constrained by historical trenches and drilling, whereas the flanks remain comparatively underexplored, motivating integration of geophysical evidence and targeting approaches.

\subsection{Mineralization types and weathering}

\begin{figure}[!htbp]
    \centering
    \begin{subfigure}[t]{0.48\textwidth}
        \centering
        \includegraphics[height=0.956\linewidth]{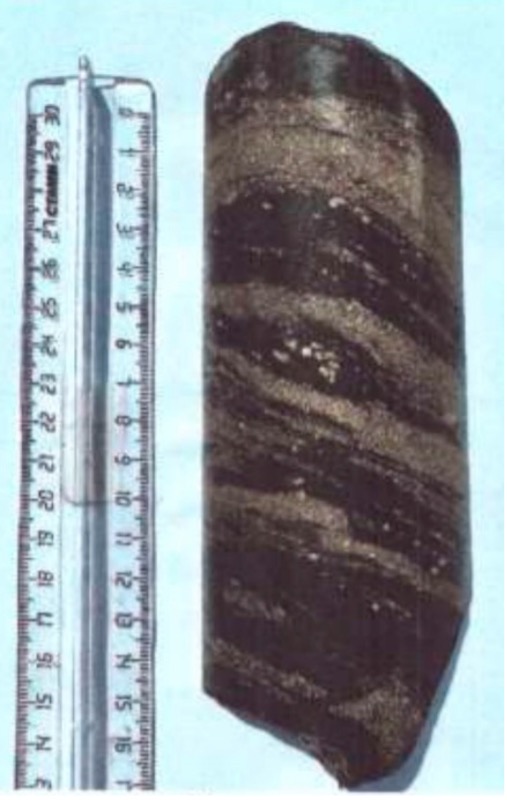}
    \end{subfigure}\hfill
    \begin{subfigure}[t]{0.48\textwidth}
        \centering
        \includegraphics[width=\linewidth]{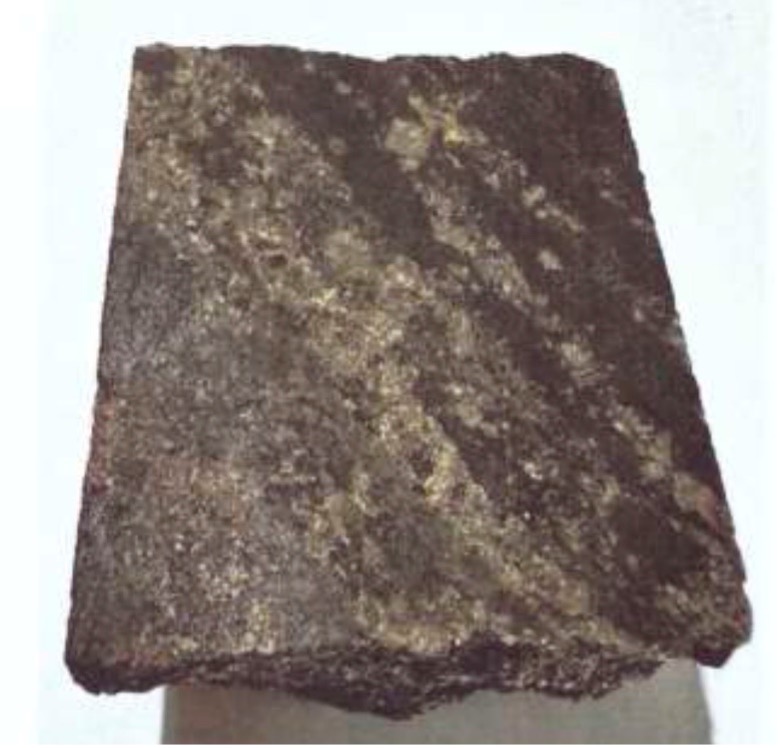}
    \end{subfigure}
    \caption{Mineralization types at Kogodai: (left) rhythmically layered pyrite mineralization within chloritolites; (right) layered pyrite mineralization with overprinting sphalerite--tetrahedrite--chalcopyrite mineralization.}
    \label{fig:mineralization_types_weathering}
\end{figure}

Within the Kogodai area, two principal mineralization styles have been distinguished (Figure~\ref{fig:mineralization_types_weathering}), together with a minor late type. The dominant style is a pyrite-rich, layered sulfur--pyrite mineralization, where pyrite is associated with quartz and chlorite, with accessory rutile/titanomagnetite and minor sphalerite; host rocks are strongly altered to Mg-rich chlorite assemblages. A second, economically more important style is a disseminated nest-like to veinlet/disseminated Zn--Cu mineralization that is spatially superimposed on, and locally overprints, the pyrite-rich mineralization. In this style, chalcopyrite and sphalerite occur both within pyrite layers and along their margins, while preserving an overall layered appearance. A third, relatively rare and late style consists of nesty disseminated copper mineralization in quartz veins, interpreted as a late-stage overprint with limited impact on overall mineralization patterns.

Near-surface weathering has produced an oxidation zone where sulfides are transformed into limonite--quartz assemblages with malachite and leached cavities; the typical depth of oxidation is reported to be on the order of $\sim$15~m, locally deeper in structurally disturbed zones. The boundaries of mineralized bodies are commonly diffuse and, given the nesty/veinlet nature of mineralization, are primarily constrained by sampling rather than by continuous visual tracing. A practical threshold grade of 0.08\% Cu has been used in prior work to outline mineralized envelopes, with higher-grade subdomains ($\geq$0.15\% Cu) used to delineate ore lenses and bodies in resource-oriented interpretations.

\section{Methods}
\label{sec3}

\subsection{Workflow}
\label{sec:methods_workflow}

\begin{figure}[!htbp]
    \centering
    \includegraphics[width=0.85\textwidth]{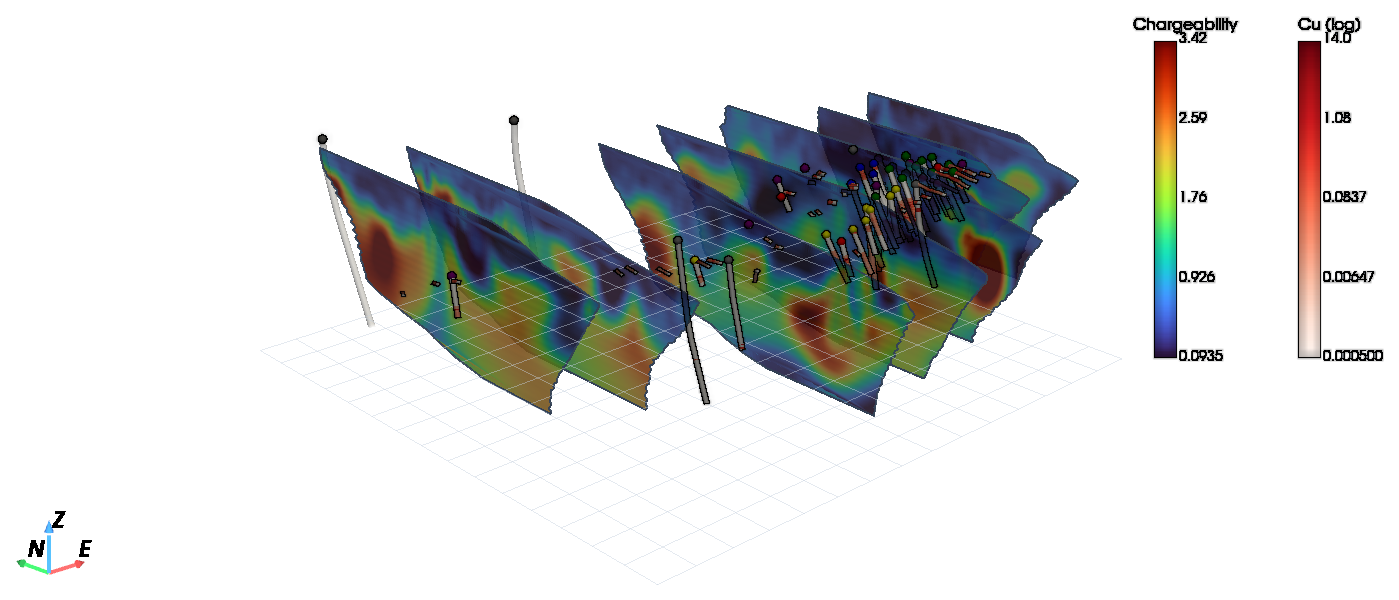}
    \caption{Three-dimensional integration of drillhole geochemistry and chargeability profiles at Kogodai, shown in oblique SW--NE view. Vertical coloured sheets are interpolated chargeability sections (seven profiles, separated along easting); thin tubes are drillholes coloured by Cu assay (log scale).}
    \label{fig:kogodai_3d_integration}
\end{figure}

\paragraph{Data available}\mbox{}\\
The proposed workflow integrates two complementary sources of information: direct geochemical observations from drillholes and trenches, and indirect geophysical observations from induced polarization (IP) and electrical resistivity surveys. Drillhole and trench assays provide direct measurements of copper grade along narrow sampling trajectories, but they only constrain a small fraction of the subsurface volume. By contrast, IP and resistivity profiles provide spatially more continuous subsurface information along 2D geophysical sections (Figure~\ref{fig:kogodai_3d_integration}). In the Kogodai dataset, these electrical datasets represent the main spatially extensive constraint on the subsurface between and below the sampled drillholes and trenches.

The use of IP and resistivity is geologically motivated by the massive-sulfide character of the Kogodai mineralization. Electrical methods, and IP in particular, are widely used in sulfide-related exploration because chargeability and resistivity variations can reflect sulfide abundance, alteration intensity, fluid pathways and lithological contrasts. Applied ER--IP studies have shown that combined resistivity and chargeability imaging can delineate mineralized, altered and oxidized zones and can support follow-up drilling, commonly with validation from borehole observations \citep{casulla2023characterization,ali2020geoelectrical,ali2020borehole}. In copper exploration, several studies have linked IP/resistivity responses to drilling-derived copper grades and used these relationships for 3D orebody modelling, geostatistical interpolation and drillhole optimization \citep{mostafaei20183d,mostafaei2019investigating,mostafaei2019mineral}. At the same time, geophysical inversions are ill-posed and non-unique; therefore, geoelectrical anomalies may be sensitive to survey geometry, regularization and data coverage. This motivates a workflow in which electrical datasets are integrated with direct assays and interpreted together with uncertainty estimates rather than used as deterministic orebody boundaries \citep{ellis2018exploring,gong2019three,su2023combining,su20253}.

\paragraph{Joint modelling}\mbox{}\\
A central practical difficulty is that the different observations are not co-located and do not have the same spatial support. Drillholes and trenches are effectively 1D sampling paths embedded in 3D space, whereas IP and resistivity data are acquired along profiles and represented as 2D sections or inversion-derived products within the same 3D coordinate system. Even where a drillhole is close to a geophysical profile, the two datasets do not sample exactly the same physical volume. Therefore, the modelling task is not a simple regression from geophysical values to copper grades at identical locations. Instead, we formulate it as the joint learning of spatially continuous 3D fields from sparse, heterogeneous and unevenly distributed observations.

Before modelling, all datasets were transformed to a common 3D coordinate frame. Drillhole and trench assays were composited to a consistent downhole support, while Cu grades, chargeability and apparent resistivity were transformed and normalized for joint training. Geophysical profile or inversion products were represented as spatial observations with associated 3D coordinates.

Recent studies demonstrate the value of combining geophysical data and machine learning for mineral prospectivity and grade prediction, including IP-based ore modelling and drilling optimization \citep{mostafaei2019investigating} and ANN-based copper-grade prediction from limited samples \citep{tsae2021ann}. Probabilistic 3D integration frameworks, including weights-of-evidence approaches, further highlight the value of combining heterogeneous datasets in a unified predictive space \citep{farahbakhsh2020weights}. Building on this line of work, we formulate the Kogodai problem as learning spatially continuous 3D fields of copper grade and geophysical attributes from incomplete observations, while explicitly quantifying epistemic uncertainty due to limited data coverage \citep{he2025survey}.

Let \(X \subset \mathbb{R}^3\) denote the spatial domain of the study volume, and let \(x_i=(x_i,y_i,z_i)\) be a sampled location. The observed target values are available only for subsets of locations. In our case, the targets are copper grade, IP chargeability and apparent resistivity. We therefore seek to learn a mapping
\begin{equation}
F_{\theta}: \mathbb{R}^{3} \rightarrow \mathbb{R}^{3}, \qquad
F_{\theta}(x, y, z)=\left(\widehat{\mathrm{Cu}}(x,y,z), \widehat{\mathrm{IP}}(x,y,z), \widehat{R}(x,y,z)\right),
\label{eq:workflow_mapping}
\end{equation}
where \(\widehat{\mathrm{Cu}}\), \(\widehat{\mathrm{IP}}\) and \(\widehat{R}\) are the predicted copper grade, chargeability and apparent resistivity, respectively. Because the three targets are not observed at all locations, training uses a masked multi-task loss: for each target, predictions are compared with ground-truth observations only at locations where that target is available. This allows drillhole/trench assays and geophysical profile observations to contribute jointly to the same continuous 3D model without requiring strict co-location.

\paragraph{Solution structure}\mbox{}\\
The field \(F_{\theta}\) is represented by a Bayesian variant of a deep fully connected neural network with a gating mechanism \citep{wang2020understandingmitigatinggradientpathologies}. A shared neural-network trunk learns a latent representation of the 3D system from spatial coordinates, while task-specific output heads predict copper grade, chargeability and resistivity. The shared trunk allows the model to learn implicit relationships between the three fields; for example, zones of elevated chargeability or low resistivity may help constrain copper prediction where direct assays are sparse \citep{nielsen2019tampia, mostafaei2019investigating}. These relationships are not imposed as hard geological rules, but are inferred from the joint spatial structure of the available observations.

The Bayesian formulation treats the network parameters as distributions rather than single deterministic values. Consequently, repeated Monte Carlo forward passes through the trained model produce multiple plausible realizations of the predicted fields. For each location in the study volume, the predictive mean is used as the estimated field value, while the variance across Monte Carlo samples is interpreted as epistemic uncertainty. This uncertainty reflects limited knowledge caused by sparse drilling, incomplete geophysical coverage and non-unique model parameterizations consistent with the data.

After training, the continuous model is sampled on the nodes of a rectilinear 3D grid to produce block-model-style outputs for copper grade, chargeability, resistivity and their associated uncertainties. These gridded outputs are visualized together with drillholes, trenches, geophysical sections and legacy geological interpretations using a PyVista-based 3D workflow. The interpretation focuses on two questions: first, whether the model recovers known mineralized zones and established geological trends; and second, whether predicted high-copper or high-chargeability zones outside the best-drilled areas define geologically plausible targets for follow-up exploration. The resulting model is therefore used as an uncertainty-aware drill-targeting tool rather than as a formal mineral resource estimate.

\subsection{Algorithm: Bayesian deep learning modelling}
\label{sec:methods_modelling}

\begin{figure}[!htbp]
    \centering
    \includegraphics[width=\textwidth]{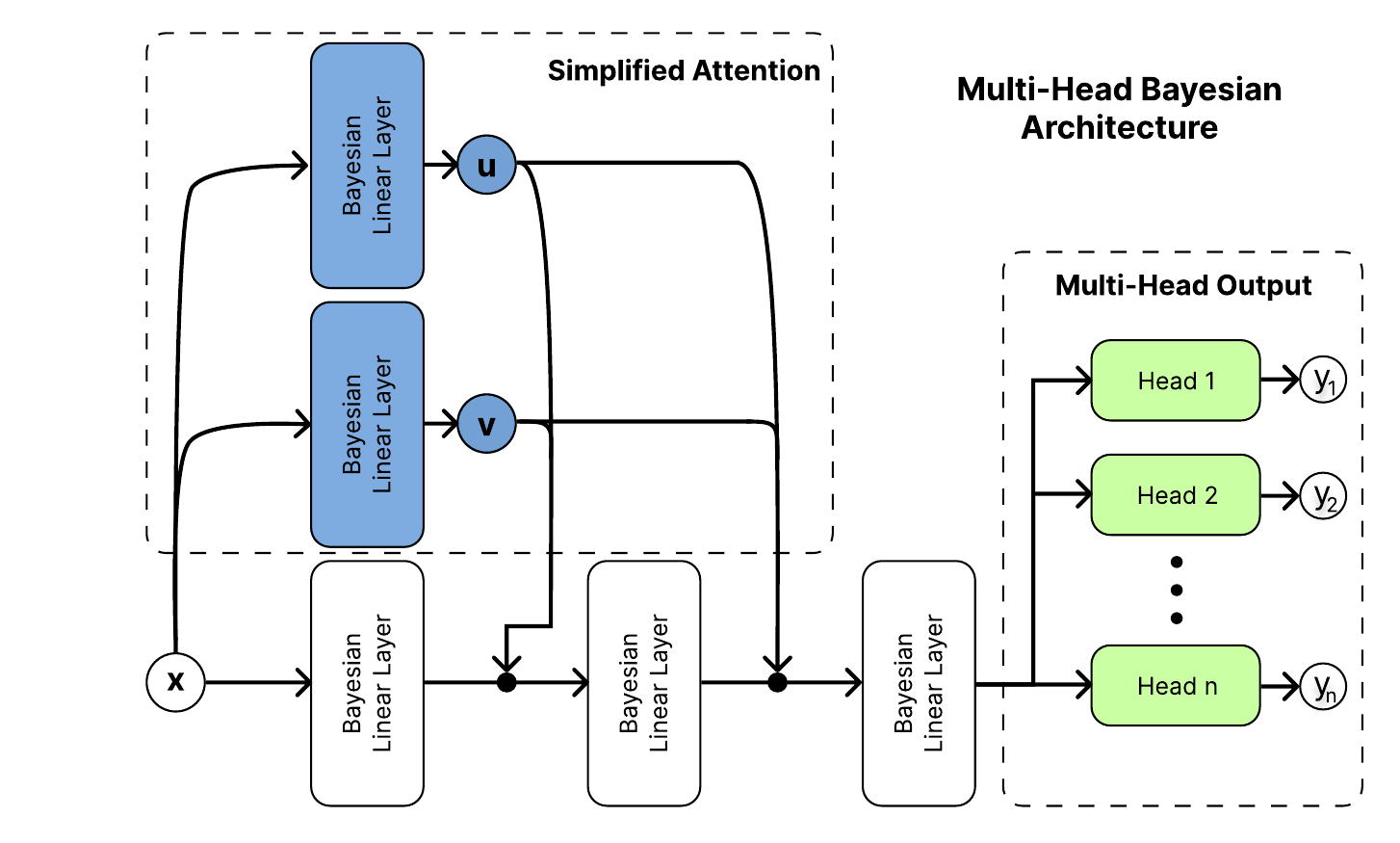}
    \caption{Architecture of the Neural Network Model. Main Part of the model is shared Modified MLP architecture containing two additional Linear Layers with activations for gating the forward path of the model and providing the latent vector. The additional set of the Head Classical MLPs are presented for each output dimension}
    \label{fig:bnn_architecture}
\end{figure}

Given the coordinate-target pairs defined in Section~\ref{sec:methods_workflow}, the modelling stage treats the continuous field as a Bayesian neural network rather than as a deterministic interpolator. Unlike deterministic deep networks that return point estimates optimized under a loss \(L(y,\hat{y})\), Bayesian neural networks place a prior over weights \(p(\theta)\) and infer a posterior distribution conditioned on the training data \((X,Y)\) \citep{he2025survey}:
\begin{equation}
p(\theta \mid X,Y) = \frac{p(Y \mid \theta, X)\,p(\theta)}{p(Y \mid X)}.
\label{eq:posterior}
\end{equation}
Different weight samples \(\theta_i\) correspond to different plausible models consistent with the observations and enable epistemic uncertainty-aware prediction.

We adopt a factorized Gaussian prior for each weight and bias parameter. For the weights of a layer with input dimensionality $d_{\mathrm{in}}$,
\begin{equation}
p(w_{ij}) = \mathcal{N}\!\left(0,\; d_{\mathrm{in}}^{-1}\right),
\label{eq:prior_weights}
\end{equation}
so that $\sigma_{\mathrm{prior}} = d_{\mathrm{in}}^{-1/2}$, analogous to He initialization scaling. For biases, a unit-variance prior $p(b_j) = \mathcal{N}(0, 1)$ is used. The approximate posterior is also factorized Gaussian,
\begin{equation}
q(w_{ij}) = \mathcal{N}\!\left(\mu_{ij},\; \sigma_{ij}^{2}\right),
\label{eq:posterior_weights}
\end{equation}
where $\mu_{ij}$ and $\log\sigma_{ij}$ are learnable parameters initialized near the prior values.

The Kullback--Leibler divergence between the posterior and prior for a single layer is computed analytically as the sum over all weight and bias parameters:
\begin{equation}
\mathrm{KL}\bigl(q(\mathbf{w})\,\|\,p(\mathbf{w})\bigr) = \sum_{ij}\left[\log\frac{\sigma_{\mathrm{prior}}}{\sigma_{ij}} + \frac{\sigma_{ij}^{2} + (\mu_{ij} - \mu_{\mathrm{prior}})^{2}}{2\,\sigma_{\mathrm{prior}}^{2}} - \frac{1}{2}\right],
\label{eq:kl_divergence}
\end{equation}
with an analogous term for biases. The total network KL is $\mathrm{KL}_{\mathrm{total}} = \mathrm{KL}_{\mathrm{backbone}} + \sum_{t}\mathrm{KL}_{\mathrm{head},t}$, summed over the shared trunk and all task-specific heads.

\paragraph{Architecture}\mbox{}\\
Our primary model is a Bayesian variant of a deep fully connected network with a gating mechanism inspired by \citet{wang2020understandingmitigatinggradientpathologies}, referred to as a Modified MLP. All linear layers are Bayesian, with learnable parameters $(\mu, \log\sigma)$ describing a factorized Gaussian posterior for each weight. To jointly predict Cu, chargeability, and apparent resistivity from sparse, non-co-located datasets, we use a multi-head design: a shared Modified MLP trunk learns a compact latent representation of the 3D system, while task-specific Bayesian MLP heads map this representation to each target variable. Head outputs are bounded by a $\tanh$ activation, consistent with target values scaled to $[-1, 1]$. This structure supports multi-task learning with heterogeneous label availability and helps maintain generalization for targets that are underrepresented in parts of the volume. The scheme of the neural-network architecture is shown in Figure~\ref{fig:bnn_architecture}.

The gating mechanism computes two modulation vectors $\mathbf{U}$ and $\mathbf{V}$ from the raw input coordinates $\mathbf{x} = (x, y, z) \in \mathbb{R}^{3}$ via two additional Bayesian linear layers:
\begin{equation}
\mathbf{U} = \sigma_{g}(\mathbf{W}_{U}\,\mathbf{x} + \mathbf{b}_{U}), \qquad
\mathbf{V} = \sigma_{g}(\mathbf{W}_{V}\,\mathbf{x} + \mathbf{b}_{V}),
\label{eq:gating_uv}
\end{equation}
where $\sigma_{g} = \mathrm{GELU}$ is the Gaussian Error Linear Unit activation. $\mathbf{U}$ and $\mathbf{V}$ are computed once per input and applied after every hidden-layer activation in the shared trunk. Denoting the pre-activation output of layer $k$ as $\mathbf{h}^{(k)}$, the forward pass through the trunk proceeds as follows:
\begin{align}
\mathbf{h}^{(0)} &= \mathbf{W}_{0}\,\mathbf{x} + \mathbf{b}_{0}, \notag \\
\mathbf{z}^{(k)} &= \sigma\!\left(\mathbf{h}^{(k-1)}\right), & k &= 1,\ldots, L\!-\!1, \label{eq:gating_forward} \\
\tilde{\mathbf{z}}^{(k)} &= \mathbf{U} \odot \mathbf{z}^{(k)} + \bigl(1 - \mathbf{z}^{(k)}\bigr) \odot \mathbf{V}, \notag \\
\mathbf{h}^{(k)} &= \mathbf{W}_{k}\,\tilde{\mathbf{z}}^{(k)} + \mathbf{b}_{k}, \notag \\
\mathbf{h}^{(L)} &= \mathbf{W}_{L}\,\mathbf{h}^{(L-1)} + \mathbf{b}_{L} \quad \text{(no activation, no gating)}, \notag
\end{align}
where $\sigma = \mathrm{SiLU}(x) = x\,\sigma_{\mathrm{logistic}}(x)$ (Sigmoid Linear Unit / Swish) is the activation used in all hidden layers of both the shared trunk and the task-specific heads. The final trunk layer $\mathbf{h}^{(L)}$ serves as the latent representation passed to each head. The gating provides an adaptive, input-dependent interpolation between two learned representations of the input at every layer, improving gradient flow and mitigating training pathologies in coordinate-based networks \citep{wang2020understandingmitigatinggradientpathologies}.

Table~\ref{tab:activations} summarizes the activation functions used in each component of the architecture.

\begin{table}[!htbp]
\centering
\caption{Activation functions by network component.}
\label{tab:activations}
\begin{tabular}{lll}
\hline
\textbf{Component} & \textbf{Activation} & \textbf{Definition} \\
\hline
Shared trunk hidden layers & SiLU (Swish) & $\mathrm{SiLU}(x) = x\,\sigma(x)$ \\
Gating layers $\mathbf{U}$, $\mathbf{V}$ & GELU & $\mathrm{GELU}(x) = x\,\Phi(x)$ \\
Head hidden layers & SiLU (Swish) & $\mathrm{SiLU}(x) = x\,\sigma(x)$ \\
Head output & Tanh & $\tanh(x) \in (-1, 1)$ \\
\hline
\end{tabular}
\end{table}

\paragraph{Training}\mbox{}\\
The model was optimized with a supervised loss evaluated only at locations where observations were available. For each target variable, a binary availability mask excluded missing labels from the loss, allowing copper assays, IP chargeability, and apparent resistivity observations to contribute jointly even when they were not co-located.

The reconstruction term for each target $k$ is a heteroscedastic Gaussian negative log-likelihood with a learnable aleatoric noise parameter $\sigma_k$ (one scalar per target):
\begin{equation}
\mathcal{L}_{k}^{\mathrm{recon}} = \frac{(y_k - \hat{y}_k)^{2}}{2\,\sigma_k^{2}} + \log\sigma_k.
\label{eq:recon_loss}
\end{equation}
The aleatoric noise scale $\sigma_k$ is initialized at approximately 5\% of each target's value range and is treated as a learnable parameter during training. The total training objective combines the weighted reconstruction losses with a KL regularization term:
\begin{equation}
\mathcal{L}_{\mathrm{total}} = \frac{\displaystyle\sum_{k} w_k\,\mathcal{L}_{k}^{\mathrm{recon}}}{\displaystyle\sum_{k} w_k} \;+\; \beta\,\frac{\mathrm{KL}_{\mathrm{backbone}} + \displaystyle\sum_{t}\mathrm{KL}_{\mathrm{head},t}}{N},
\label{eq:training_loss}
\end{equation}
where $w_k = n_k$ is the number of available observations for target $k$ (proportional weighting), $N = \sum_k n_k$ is the total training-set size, and $\beta$ is a fixed KL weight that is \emph{not} annealed during training, the final model uses $\beta = 0.01$. This multi-target formulation allows the shared trunk to learn spatial relationships among the fields while respecting the heterogeneous coverage of the input datasets.

The neural-network configuration selected is:
\begin{itemize}
    \item Shared trunk: 4 hidden Bayesian linear layers with hidden dimension 256, two gating layers, and a latent output dimension of 16;
    \item Task-specific heads: 2 Bayesian linear layers with hidden dimension 256 for each of the four targets (resistivity, chargeability, chargeability from 3D voxel model, Cu).
\end{itemize}

Optimization uses AdamW with $\beta_{\mathrm{Adam}} = (0.9, 0.99)$, weight decay $= 0$, batch size 64, and cosine learning-rate decay from a peak of $10^{-3}$ after a 50-epoch linear warmup, over a total of 550 epochs (50 warmup + 500 decay).

Training batches are formed \emph{separately} for each target: four independent data loaders (one per target, each shuffled) are iterated in every training step. The number of steps per epoch is determined by the shortest loader (Cu, $\approx$38 steps at batch size 64), so that approximately 100\% of Cu data and $\approx$32\% of geophysical data are seen per epoch; exhausted loaders are cycled.

\paragraph{Data preprocessing}\mbox{}\\
The drillholes data was sampled by intervals with step 10 meters and interpolated with PCHIP method to the 2 meters grids.

As part of data preprocessing, coordinate features were scaled to $[-1, 1]$. Target geophysical values and Cu measurements were log-transformed and then scaled to $[-1, 1]$ using a min--max scaler. Because the scaled targets span $[-1, 1]$, the head output layers use a $\tanh$ activation that constrains predictions to the same range.

\paragraph{Monte Carlo sampling}\mbox{}\\
Epistemic uncertainty was quantified via Monte Carlo (MC) sampling from the learned weight distributions. During training, $R = 16$ independent weight samples are drawn per forward pass; for calibration evaluation, $R = 256$ samples are used to obtain stable uncertainty estimates. For a given input location \(\mathbf{x}\), repeated forward passes yield predictions \(y_r = F(\mathbf{x};\theta_r)\), $r = 1,\ldots,R$. The predictive mean and variance are estimated as:
\begin{equation}
\hat{y} = \frac{1}{R}\sum_{r=1}^{R} y_r,
\label{eq:mc_mean}
\end{equation}
\begin{equation}
\sigma_y^2 = \frac{1}{R-1}\sum_{r=1}^{R}\left(y_r-\hat{y}\right)^2,
\label{eq:mc_var}
\end{equation}
where $\theta_r \sim q(\theta)$ are independently sampled weight configurations. The resulting uncertainty maps were interpreted jointly with mean predictions to support drill-targeting decisions by distinguishing areas of strong support (low uncertainty) from areas where additional drilling would be most informative (high epistemic uncertainty).

\section{Induced Polarization and Resistivity data}
\label{sec:geophysical_data_preparation}

\begin{figure}[!htbp]
    \centering
    \begin{subfigure}[c]{0.32\linewidth}
        \centering
        \includegraphics[width=\linewidth]{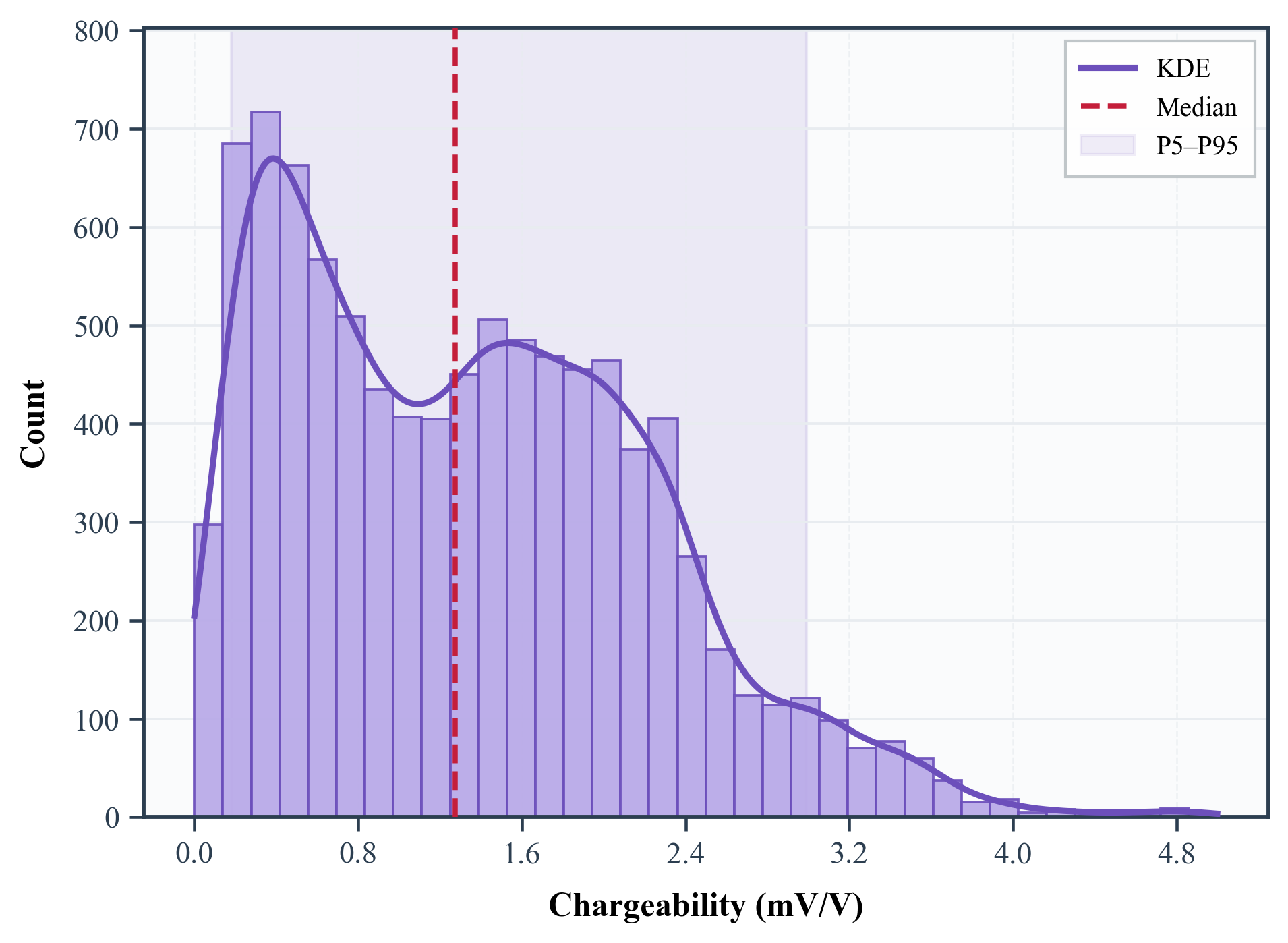}
    \end{subfigure}\hfill
    \begin{subfigure}[c]{0.32\linewidth}
        \centering
        \includegraphics[width=\linewidth]{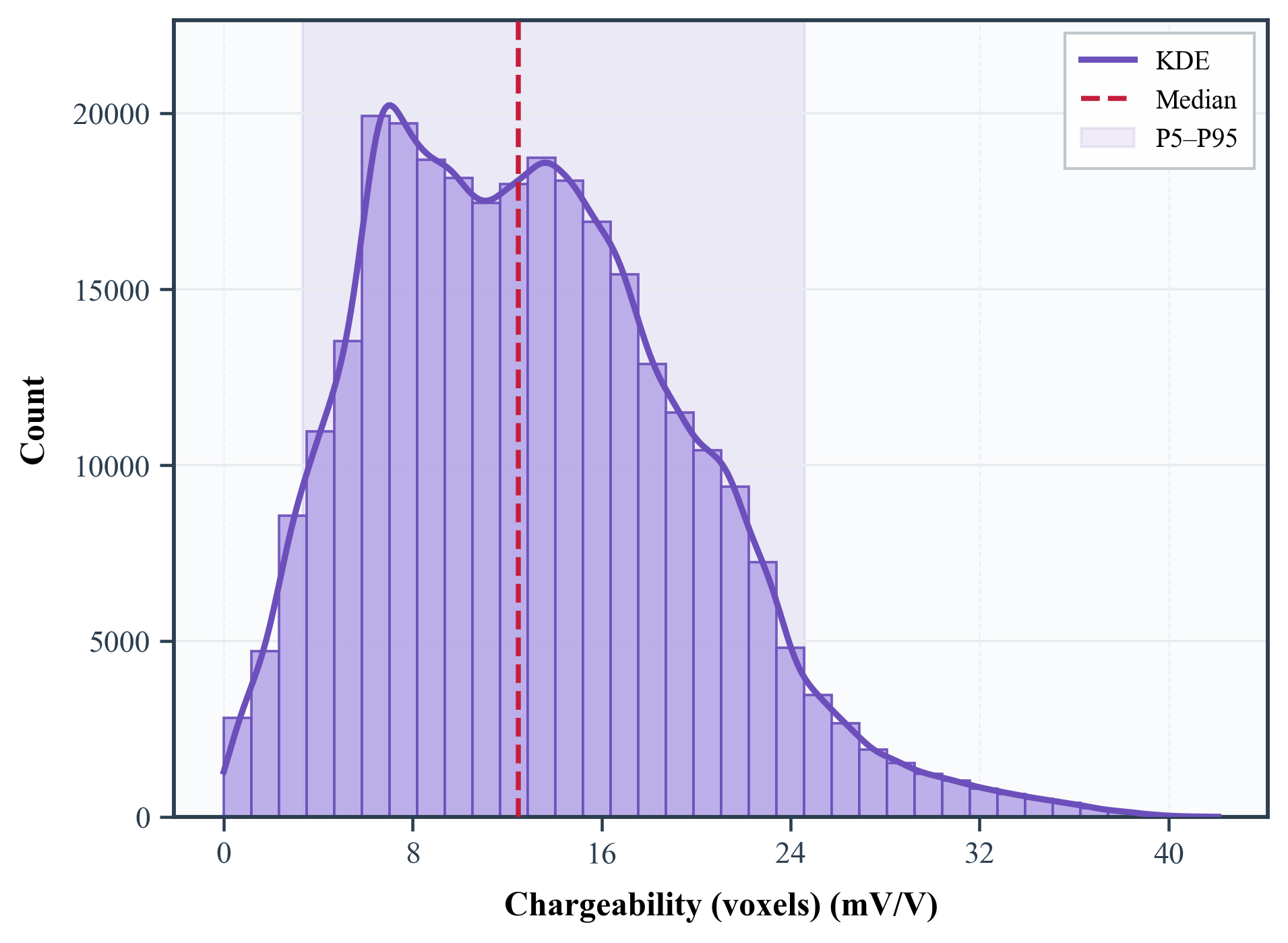}
    \end{subfigure}\hfill
    \begin{subfigure}[c]{0.32\linewidth}
        \centering
        \includegraphics[width=\linewidth]{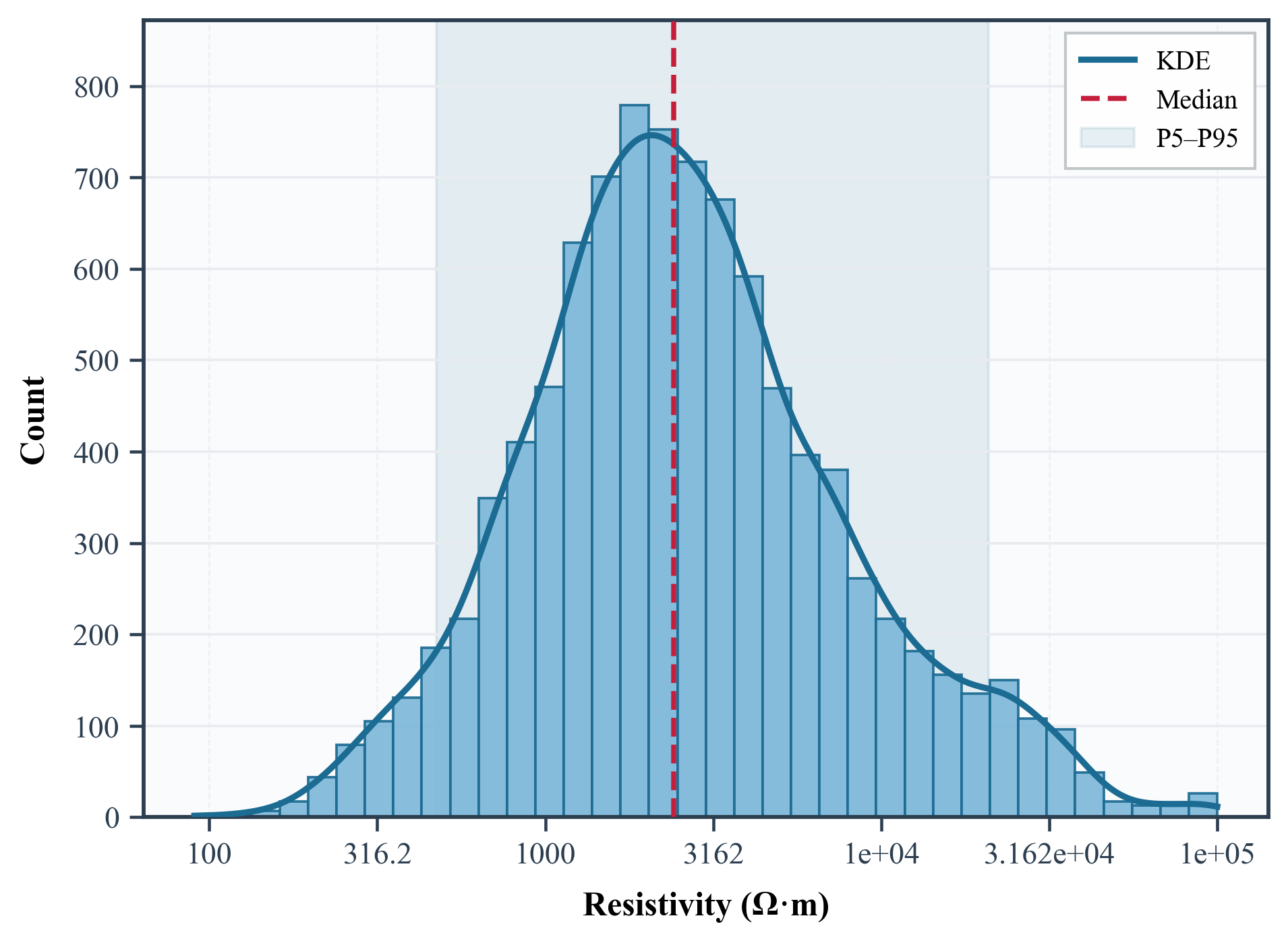}
    \end{subfigure}
    \caption{Distributions of Kogodai IP chargeability and apparent resistivity data used for geophysical data preparation and modelling. Left: histogram of chargeability values from Kogodai IP profile sections; the distribution of 9,503 positive samples is shown on a linear scale (mV/V), with the purple kernel density estimate (KDE), red dashed median (1.27~mV/V), and 5th--95th percentile band (0.18--2.99~mV/V). Centre: histogram of chargeability from the 3D IP inversion voxel model; the distribution of 292,677 positive, finite voxels is shown on a linear scale (mV/V), with KDE, median (12.5~mV/V), and P5--P95 band (3.3--24.6~mV/V). Right: histogram of resistivity values from Kogodai IP profile sections; values are log$_{10}$-transformed before binning so bins are equal width in log space, the x-axis is labelled in $\Omega\cdot$m, and the 9,535 positive samples span approximately 90--100,000~$\Omega\cdot$m, with KDE, median (2,410~$\Omega\cdot$m), and P5--P95 shading (473--20,900~$\Omega\cdot$m) computed in log$_{10}$ space.}
    \label{fig:ip_resistivity_histograms}
\end{figure}

Geophysical data provide the main source of spatially continuous subsurface information outside the footprint of existing drilling and trenching. In this study, the objective was not to refine the detailed geometry of known ore lenses, but to identify and prioritize prospective locations for follow-up drilling where direct sampling is absent. Consequently, \textit{induced polarization (IP) chargeability} and \textit{apparent resistivity} were treated as primary constraints for 3D targeting (Figure~\ref{fig:ip_resistivity_histograms}).

\subsection{Acquisition and quality control}
\label{sec:geophys_acquisition_qc}

Ground electrical surveys were carried out by Aurora Geophysics Limited using induced-polarization electrical tomography with a modern GDD Instrumentation acquisition system. Measurements were acquired along profiles using a \textit{pole--dipole array}, selected as a compromise between depth of investigation and field productivity in the Kogodai area. The basic electrode spacing, including the receiver-dipole length and current-electrode step, was 25~m. The spacing between profiles was either 200 or 400~m. Chargeability was recorded for several time windows after current shut-off and as an averaged value.

Initial quality control and rejection of noisy measurements were performed by Aurora Geophysics Limited in \textit{Oasis Montaj} using the IP module. Measurements were inspected in tabular and graphical form, including \textit{pseudo-sections} of apparent resistivity and IP parameters. The \textit{average off-time chargeability} was retained as the principal IP interpretation parameter because it is less sensitive to transient noise than individual time-window measurements.

\subsection{Inversion workflow}
\label{sec:geophys_inversion}

Further processing and inversion were performed in \textit{ZondRes2D}. Chargeability values were converted from mV/V to percent before integration. ZondRes2D performs \textit{2.5D inversion} of resistivity and IP profile data, assuming that the model is two-dimensional in the section plane and laterally invariant perpendicular to the profile. The inversion mesh used a regular grid with 20 vertical layers of increasing thickness, and the model depth reached approximately 350~m on the Kogodai profiles.

Inversion was carried out iteratively using a Newton or quasi-Newton optimization scheme with regularization. The resistivity section was estimated first, followed by the IP chargeability section. Among the available inversion options, the \textit{Occam algorithm} was used because it provides comparatively smooth and stable solutions. The resulting sections were therefore interpreted as \textit{model-dependent geophysical evidence}, not as unique geological boundaries.

\bigskip
The available IP/resistivity archive contained several derived products in heterogeneous formats including \textit{3D chargeability voxel model}, and gridded horizon-wise \textit{level slices} for both chargeability and apparent resistivity. Such heterogeneity is common in legacy exploration archives.

Because provenance metadata were incomplete and anomaly expression differed across products, we tested model training against both horizon-wise slices and the 3D chargeability voxel representation. For chargeability, the final input retained values from the layers. Obvious non-physical values were excluded as artefacts.

\section{Drilling and trench sampling data}
\label{sec:drilling_trench_sampling_data}

Exploration drilling and trenching at Kogodai were carried out intermittently between 2014 and 2024. The recent exploration programme completed \textit{4,949 linear metres of drilling (45 drillholes) and 1,207 linear metres of trenches (35 trenches)}. Together with legacy records, these data provide the direct sampling basis for the copper assay dataset used in this study.

Core was not sampled continuously along the full drillhole length. Sampling focused on intervals with visible sulfide mineralization, with at least two bounding or shoulder samples collected before and after each mineralized zone. Core-sample lengths ranged from 0.3 to 2.0~m, with an average length of approximately 1.0~m. This selective sampling strategy is important for modelling because unsampled intervals do not represent missing-at-random data, but commonly correspond to intervals without recognized visible mineralization. For modelling purposes, these unsampled regions were assigned copper values at the analytical detection limit.

 Of the 7,136~m of total core recorded in the database, 1,876~m were assayed, corresponding to 26.3\% sampled core and 73.7\% unsampled core. Median per-hole assay coverage is 36.0\%, and 37 of 44 drillholes contain assays.

Drilling and trench datasets were compiled to provide 3D constraints on copper mineralization and to enable joint analysis with the IP/resistivity products. The available information included collar coordinates, downhole survey measurements (inclinometry), trench picket coordinates and geochemical assay intervals. Drillhole trajectories were reconstructed with a \textit{desurvey workflow} in Python using WellPathPy\footnote{\url{https://wellpathpy.readthedocs.io/}.}, which converts collar coordinates and survey angles into a polyline representation. For trenches with multiple picket points, trench geometry was approximated by fitting a smoothed 3D curve through the pickets using geomdl\footnote{\url{https://nurbs-python.readthedocs.io/}.}, and sample intervals were then mapped into the same 3D coordinate framework.

\begin{figure}[!htbp]
    \centering
    \begin{subfigure}{0.49\linewidth}
        \centering
        \includegraphics[width=\linewidth]{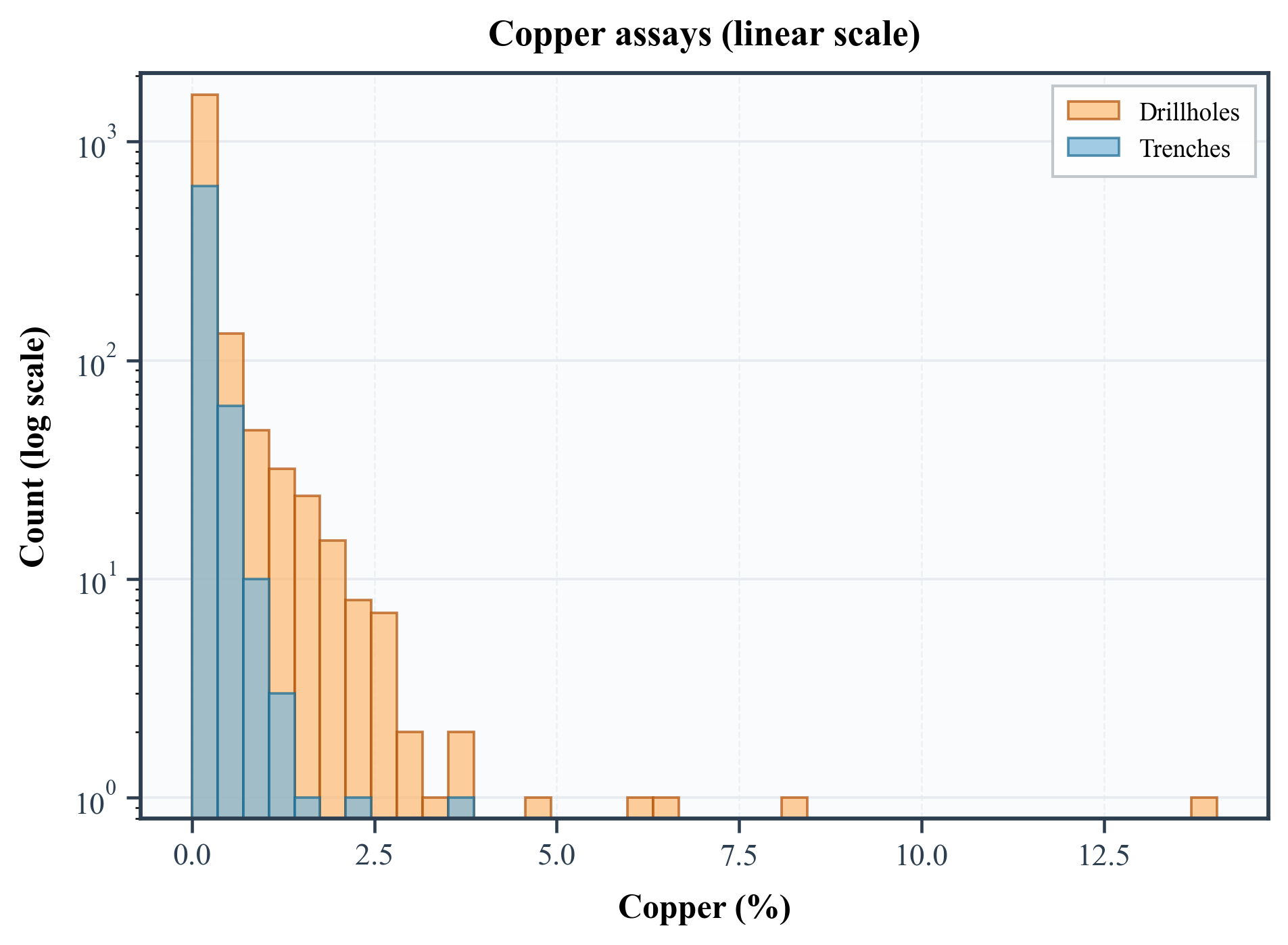}
    \end{subfigure}
    \hfill
    \begin{subfigure}{0.49\linewidth}
        \centering
        \includegraphics[width=\linewidth]{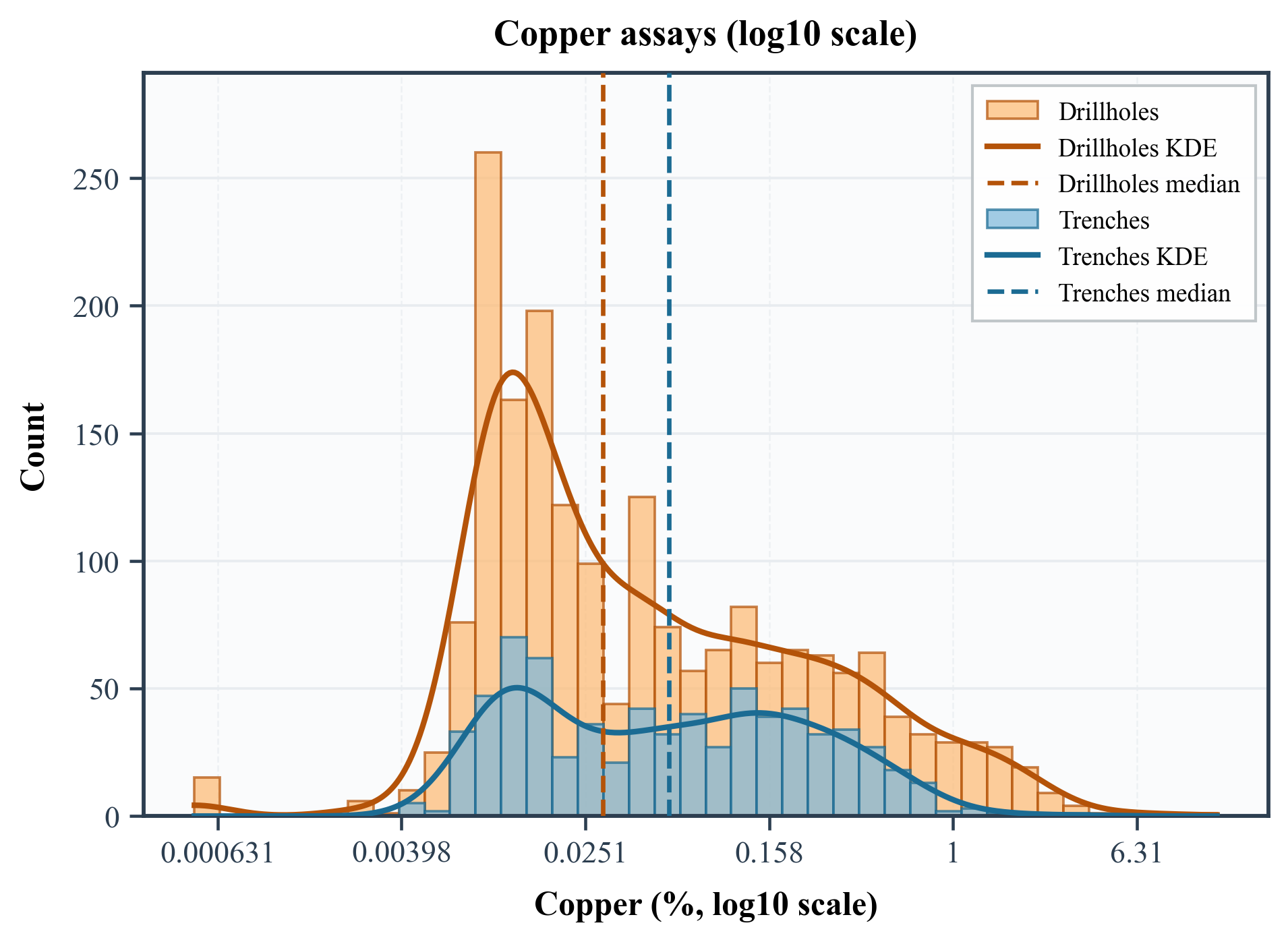}
    \end{subfigure}
    \caption{Distribution of copper grades in drillhole and trench samples: original-scale values (left) and log-transformed values (right).}
    \label{fig:cu_dist}
\end{figure}

Copper values are available for \textit{2,628 samples} in the working assay database. The distribution is strongly right-skewed. Accordingly, prior to modelling we applied a log-transform and normalization so that drilling and trench assays could be combined more robustly with the geophysical targets.

Finally, we addressed the scale mismatch between geochemical intervals and geophysical observations. IP/resistivity profiles have an effective spatial sampling on the order of \(\sim 15\)~m, whereas assay intervals are typically much shorter. To harmonize scales for machine-learning-based integration (rather than resource estimation), drillhole and trench assays were composited into \textit{10~m downhole-length intervals}. This compositing step reduces high-frequency variability, including very thin mineralized lenses, but provides a consistent support size comparable to the geophysical datasets and enables joint 3D modelling for drill-targeting analysis.

\section{Predicted 3D mineralization model}
\label{sec:results}

\begin{figure}[!htbp]
    \centering
    \includegraphics[width=0.82\textwidth]{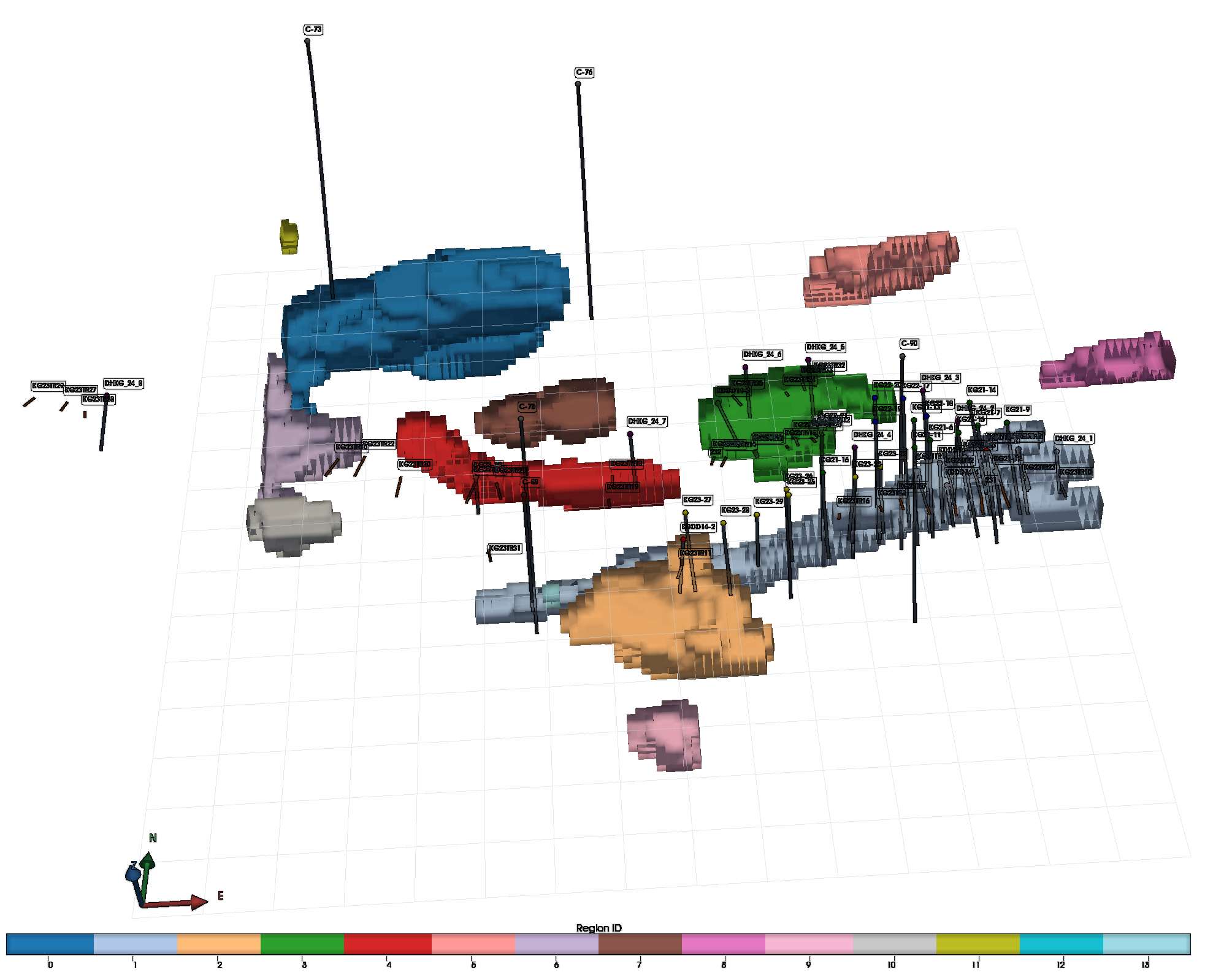}
    \caption{3D mean Cu prediction from the Bayesian Modified MLP model after thresholding at 0.15~\% Cu. Colours indicate delineated zones identified from the model output. The visualization is intended for qualitative interpretation and drill targeting; it should not be treated as a resource-grade block model.}
    \label{fig:mineralization}
\end{figure}

The Bayesian Modified MLP model produces 3D predictions within the volume constrained by the integrated dataset (chargeability, apparent resistivity, and the 3D inverted chargeability voxel model; see Section~\ref{sec:geophysical_data_preparation}). The model jointly estimates the mean Cu distribution and epistemic uncertainty; for visualization we highlight blocks exceeding 0.15~\% Cu (Figure~\ref{fig:mineralization}). The resulting geometry emphasizes a principal mineralized trend in the southern part of the area and several smaller localized features that spatially coincide with elevated IP responses, particularly in the inverted chargeability products.

\begin{figure}[!htbp]
    \centering
    \includegraphics[width=\textwidth]{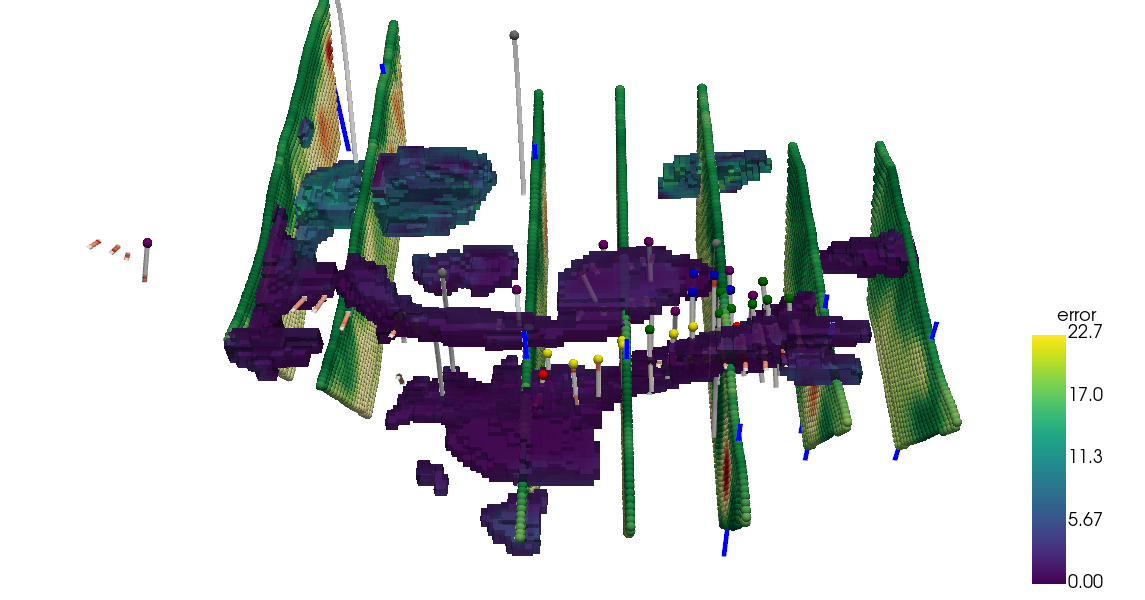}
    \caption{Relative epistemic uncertainty expressed as $3\sigma(\mathrm{Cu})/\mathrm{Cu}$ from Monte Carlo sampling of the Bayesian model (Section~\ref{sec:methods_modelling}). Darker colours indicate higher confidence (lower relative uncertainty). Uncertainty decreases in areas supported by denser sampling and increases toward poorly constrained parts of the volume.}
    \label{fig:mineralization_uncertainty}
\end{figure}

\begin{figure}[!htbp]
    \centering
        \includegraphics[width=0.7\linewidth]{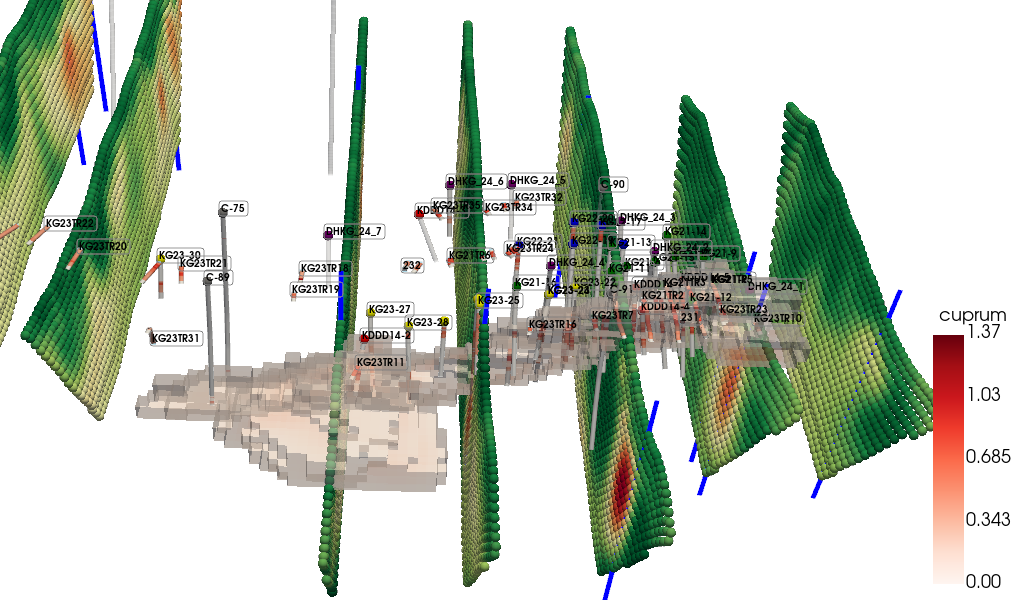}
    \caption{Example of a model-derived candidate mineralization feature (predicted Cu, \%) shown as a representative target. The feature spatially coincides with elevated IP responses and illustrates how the model highlights potential extensions beyond the immediate footprint of sampling.}
    \label{fig:example_target}
\end{figure}

At the scale of the study area, the predictions suggest a principal mineralized trend and several smaller localized features. Many of these features preferentially occur where chargeability anomalies are expressed most clearly in inversion-derived products, while uncertainty increases toward parts of the volume with sparse sampling and/or limited geophysical constraint (Figure~\ref{fig:mineralization_uncertainty}). These localized features are interpreted as candidate targets that warrant follow-up validation (Figure~\ref{fig:example_target}).

\subsection{Mineralized zones used for geological comparison}
\label{sec:mineralized_zones_comparison}

The previous expert geological interpretation of Kogodai identifies five copper-pyrite mineralized zones. These zones provide the main geological reference for evaluating the neural-network predictions. In the unpublished GEOS LLP company technical report used as the resource-reference document, individual mineralized bodies and lenses are delineated within these zones using a 0.15\% Cu threshold; however, their boundaries are diffuse and largely assay-defined. Because our workflow operates at a coarser support than the resource interpretation---including 10~m assay composites designed for integration with geophysical data---we compare the model outputs primarily with the broader mineralized zones rather than with individual resource lenses \citep{pyatkov2025kogoday}.

At the scale of the present study, these five zones (Table~\ref{tab:mineralized_zones_comparison}) are more informative than individual resource bodies because they express the main geological controls on mineralization: amphibolite-hosted or amphibolite-contact settings, subparallel east--northeast to west--northwest trends, local fault control, and variable degrees of surface and down-dip constraint. In terms of validating the current resource interpretation, the model successfully identifies the principal known mineralized zones within the area covered by the geophysical voxel data. This result is particularly important because no explicit geological interpretation, zone boundaries, or resource-model geometry were provided to the neural network as guidance. Instead, the model reconstructed the known mineralization pattern entirely from the integrated drilling, trench and geophysical data, supporting its use as an independent data-driven check on the geological interpretation. This provides confidence that AI-based targeting can be relied upon as a second opinion for validating established interpretations and highlighting targets for further review.

\begingroup
\footnotesize
\setlength{\tabcolsep}{3pt}
\renewcommand{\arraystretch}{1.22}
\begin{longtable}{>{\raggedright\arraybackslash}p{0.17\textwidth}>{\raggedright\arraybackslash}p{0.57\textwidth}>{\raggedright\arraybackslash}p{0.20\textwidth}}
\caption{Mineralized zones used as geological references for interpreting model predictions. The corresponding model zones refer to the separation of predicted zones using a 0.15\% Cu cutoff.}
\label{tab:mineralized_zones_comparison}\\
\hline
\textbf{Zone} & \textbf{Geological description} & \textbf{Corresponding zone in Figure~\ref{fig:mineralization}} \\
\hline
\endfirsthead
\hline
\textbf{Zone} & \textbf{Geological description} & \textbf{Corresponding zone in Figure~\ref{fig:mineralization}} \\
\hline
\endhead
Zone~1---Main mineralized zone & Largest, best-studied southern zone; exposed for more than 650~m with an east--northeast strike of approximately 80\(^{\circ}\) and surface width up to 70~m. Traced down-dip to about 200~m centrally, 110~m on the western flank and 20~m on the eastern flank; flanks remain open. & Zone~1 (grey). \\
Zone~2 & Approximately 150~m northwest of the Main zone; traced for about 280~m at surface, strikes approximately 75\(^{\circ}\), and is 6--10~m wide. Intersected by seven trenches and three drillholes, with down-dip continuity to about 50~m. Hosted along the amphibolite footwall--schist/gneiss hanging-wall contact; the eastern flank remains underexplored. & Zone~3 (green; southern part). \\
Zone~3 & About 110~m north of Zone~2; traced for about 160~m at surface, strikes approximately 80\(^{\circ}\), and is about 10~m wide. Intersected by four trenches and two drillholes, with down-dip continuity to about 35~m. Eastern flank follows the same amphibolite--schist/gneiss contact as Zone~2; western flank lies within amphibolite and is better constrained. & Zone~3 (green; northern part). \\
Zone~4 & Approximately 380~m west of Zone~2, on the western side of a Quaternary-filled valley, and may continue the same mineralized trend. Traced for about 190~m at surface, sub-latitudinal, up to 13~m wide, and intersected by three trenches and one drillhole; down-dip extent is about 40~m. Amphibolite-hosted with a hanging-wall fault zone marked by fracturing, cataclasis and limonitization; eastern flank remains under cover. & Zone~4 (red). \\
Zone~5 & Approximately 450~m northwest of Zone~4 and possibly its continuation. Amphibolite-hosted, traced for about 130~m, strikes approximately 110\(^{\circ}\), and is up to 12~m wide. Intersected by three trenches and one drillhole, traced down-dip to about 80~m, associated with the same hanging-wall fault structure as Zone~4, and open to the northwest and southeast. & Outside the geophysical voxel-data coverage; no prediction was made. \\
\hline
\end{longtable}
\endgroup

\subsection{Prediction quality across prospectivity thresholds}
\label{sec:prediction_quality}

The continuous Cu-grade prediction can also be converted into a binary prospectivity model by applying a cutoff to the predicted Cu field. For a given threshold, blocks with predicted Cu above the cutoff are treated as prospective, whereas blocks below the cutoff are treated as non-prospective. Varying this cutoff therefore generates a family of prospectivity maps and makes it possible to evaluate how sensitive target delineation is to the chosen prediction threshold.

\begin{figure}[!htbp]
    \centering
    \begin{subfigure}[t]{0.49\textwidth}
        \centering
        \includegraphics[width=\linewidth]{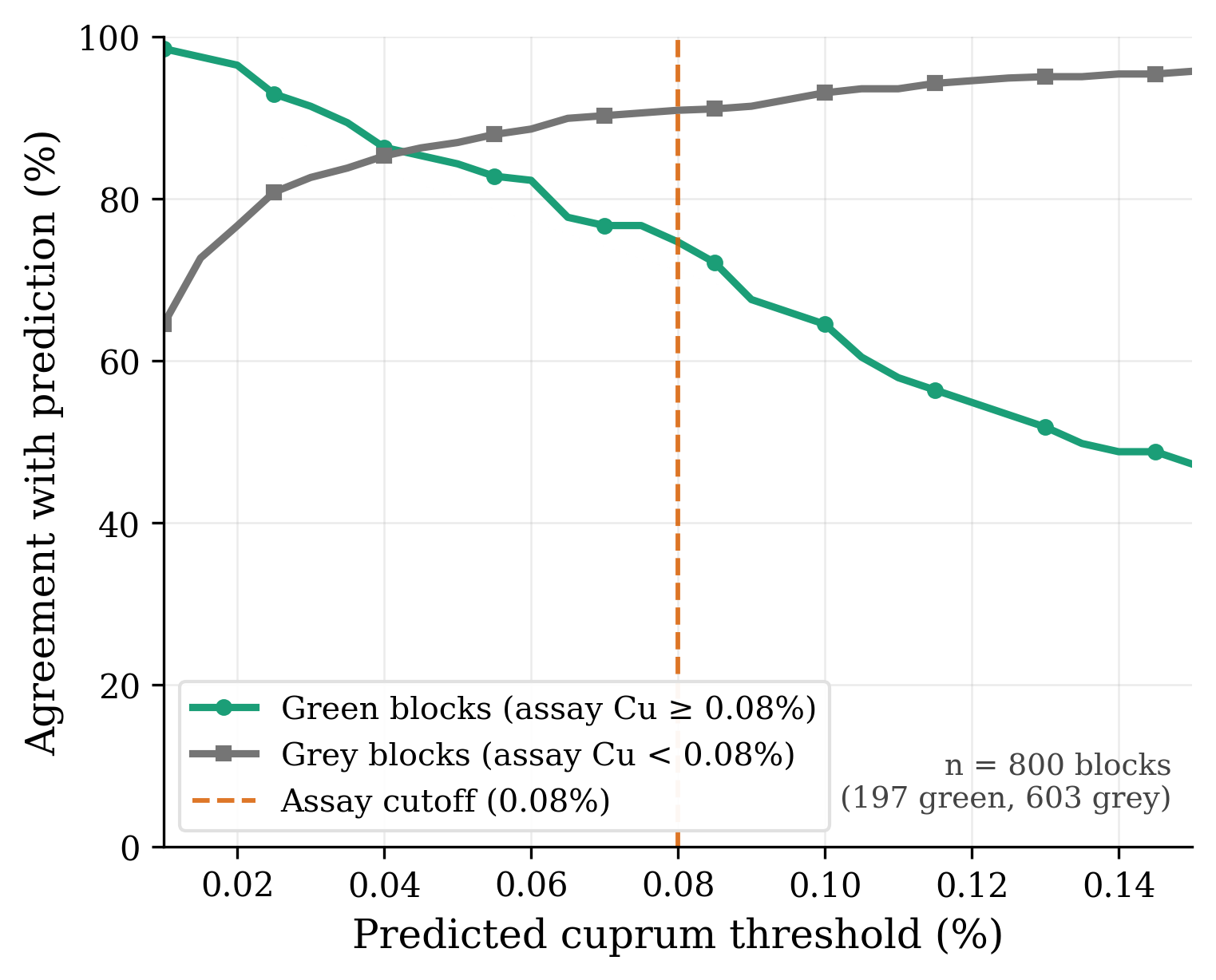}
        \caption{Threshold sensitivity.}
        \label{fig:cu_threshold_sensitivity}
    \end{subfigure}\hfill
    \begin{subfigure}[t]{0.49\textwidth}
        \centering
        \includegraphics[width=\linewidth]{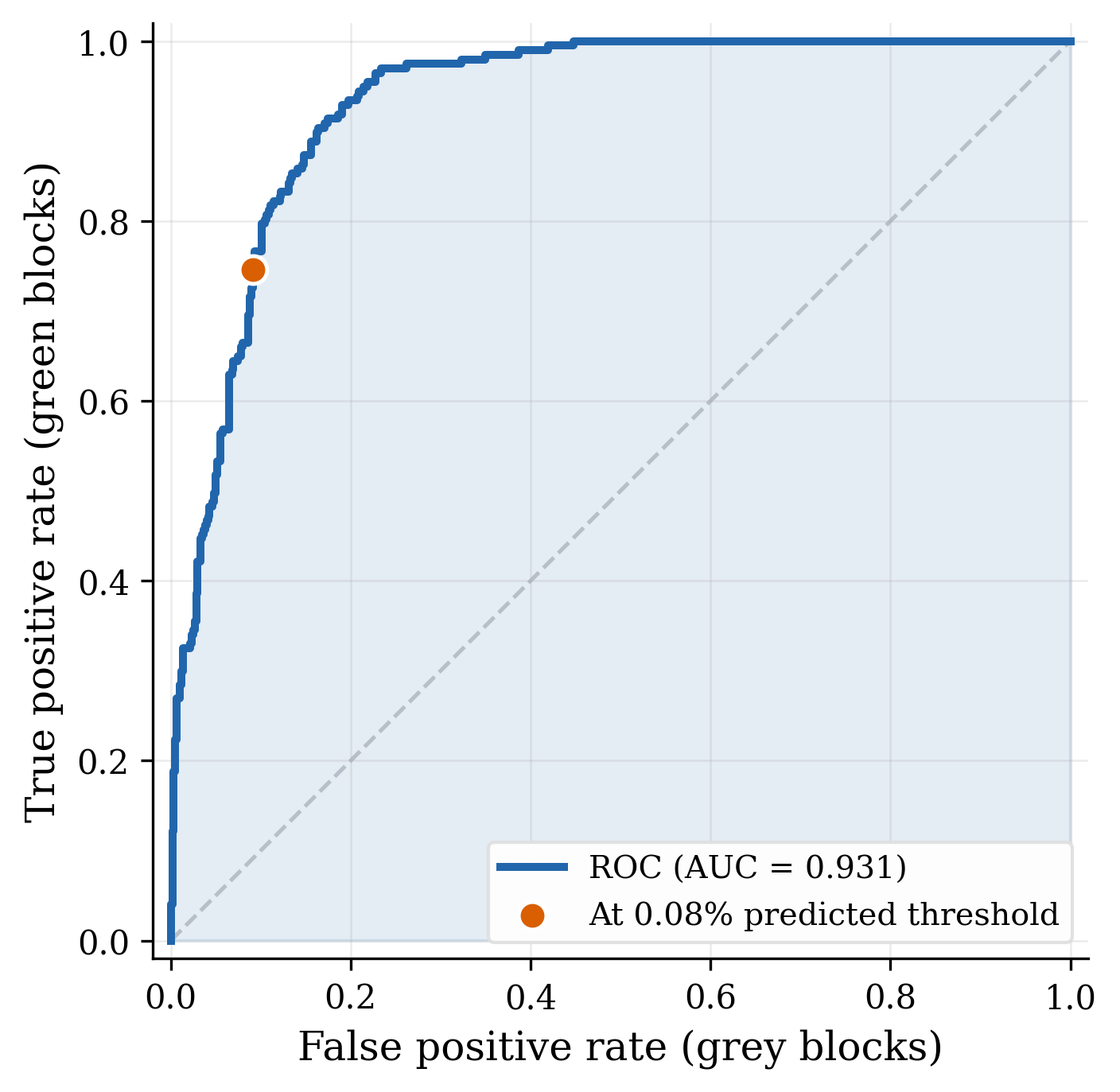}
        \caption{ROC curve.}
        \label{fig:cu_roc_curve}
    \end{subfigure}
    \caption{Prediction-quality assessment for thresholded Cu-grade predictions.}
    \label{fig:prediction_quality}
\end{figure}

We assessed prediction capability over a range of Cu thresholds by comparing predicted block values with assay-derived block labels. Sampled voxels intersecting drillholes or trenches were assigned to the ore-bearing class when the corresponding assays contained Cu values of at least 0.08~\%; all remaining sampled voxels were assigned to the ore-free class. For thresholds from 0.01 to 0.15~\% Cu, we then measured whether ore-bearing blocks were correctly marked as prospective and whether ore-free blocks were correctly marked as non-prospective.

The threshold-sensitivity curves show that the predicted Cu field retains a high fraction of ore-bearing blocks across practical cutoff values while rejecting most ore-free blocks near the 0.08~\% reference threshold (Figure~\ref{fig:prediction_quality}a). The corresponding receiver operating characteristic curve evaluates the discrimination of ore-bearing from ore-free sampled blocks using the predicted Cu field (Figure~\ref{fig:prediction_quality}b). The true positive rate is the fraction of ore-bearing blocks classified above the threshold, whereas the false positive rate is the fraction of ore-free blocks classified above the threshold. The area under the curve (AUC~=~0.93) indicates strong separation between the two classes. At a predicted Cu threshold of 0.08~\%, the model achieves a true positive rate of approximately 0.75 and a false positive rate of approximately 0.09. This supports the use of thresholded grade predictions as prospectivity maps for drill targeting, while also making the trade-off between target inclusiveness and false positives explicit.

\begin{figure}[!htbp]
    \centering
    \includegraphics[width=0.8\textwidth]{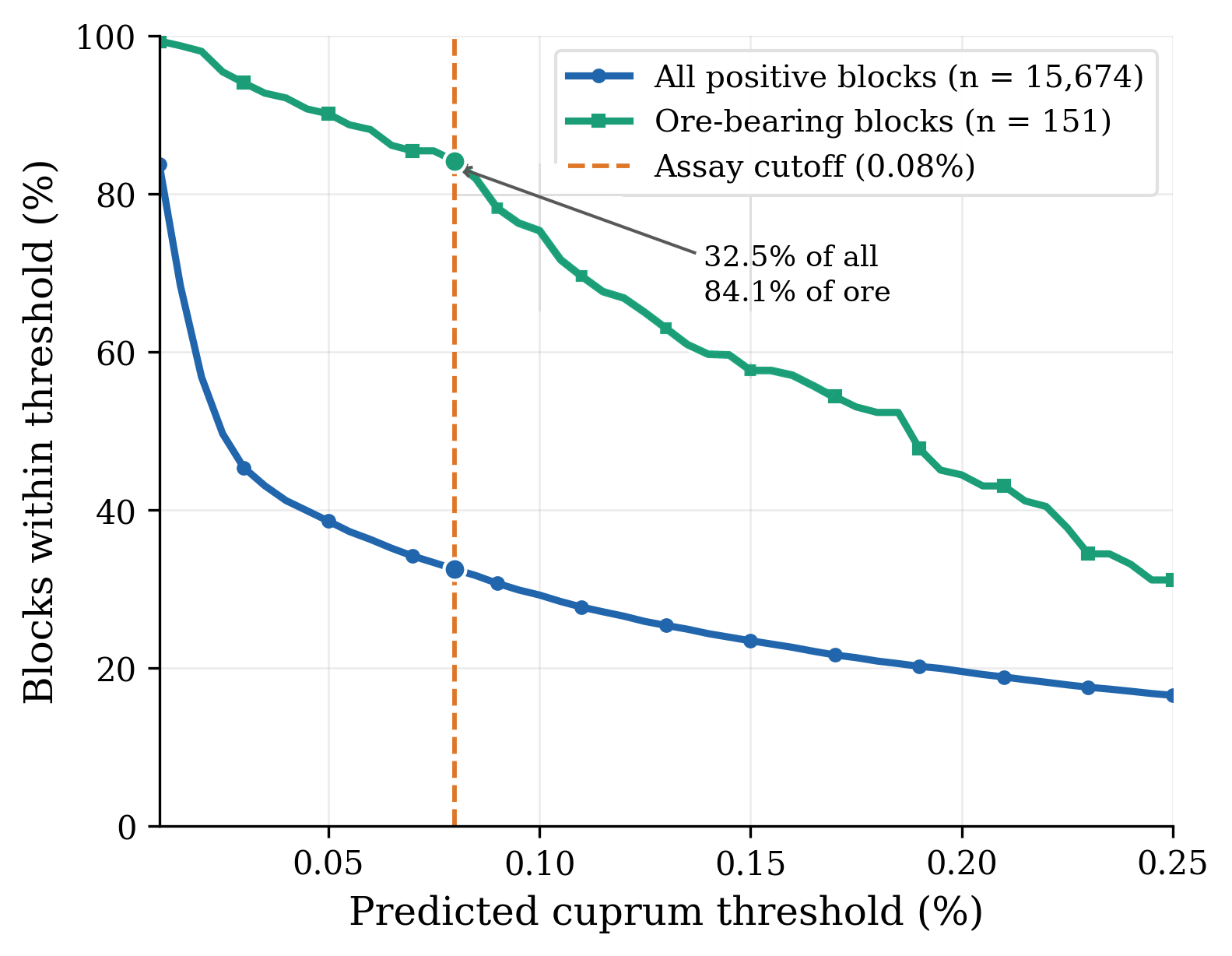}
    \caption{Ore-capture curves for confident positive predicted-Cu blocks and ore-bearing assay blocks after applying the confidence prefilter.}
    \label{fig:cu_ore_capture}
\end{figure}

Ore-capture curves provide an additional way to evaluate whether thresholded predictions retain known mineralization within the subset of blocks where the model prediction is both positive and relatively confident. The curves were computed only on blocks that pass a confidence prefilter: predicted Cu $>$ 0~\% and relative epistemic uncertainty $3\sigma(\mathrm{Cu})/\mathrm{Cu} < 1$. This retains 15,674 blocks, corresponding to 3.2~\% of all modelled blocks (484,770) and 39.5~\% of blocks with positive Cu predictions (39,662). In Figure~\ref{fig:cu_ore_capture}, the blue curve shows the fraction of these confident positive blocks below each predicted Cu threshold, whereas the green curve shows the same cumulative fraction for the 151 ore-bearing blocks (assay Cu $\geq$ 0.08~\%). At the assay cutoff of 0.08~\% Cu, 32.5~\% of all confident positive blocks and 84.1~\% of ore-bearing blocks fall within the predicted threshold, indicating that the threshold captures most known ore-bearing blocks while retaining a more selective subset of confident positive predictions.

\section{Discussion}
\label{sec:discussion}

\subsection{Model outputs as a decision-support layer for drill targeting}

This study demonstrates a pragmatic framework for drill targeting at an early-stage brownfield prospect by integrating heterogeneous information sources within a single 3D predictive model. The workflow combines direct copper observations from drillhole and trench assays with indirect geophysical constraints from induced polarization (IP) chargeability and apparent resistivity products, including profile-based sections, horizon-wise slices and a 3D chargeability voxel. The objective is not to replace geological interpretation or to produce a formal mineral resource estimate, but to provide a decision-support layer for ranking exploration hypotheses and prioritizing follow-up drilling.

A key advantage of the Bayesian formulation is that it provides not only mean predictions of copper grade and geophysical attributes, but also an estimate of epistemic uncertainty. This is particularly important at Kogodai, where direct sampling is sparse, selective and unevenly distributed, and where the geophysical information is available mainly along profiles. The resulting uncertainty field allows predicted targets to be separated into two practical categories. The first consists of zones where elevated predicted copper is supported by multiple, partially independent constraints, such as assays and geophysical responses, and where uncertainty is relatively low. These areas represent more robust targets within the current evidence base. The second consists of zones where elevated predictions are driven mainly by one data source or occur in poorly sampled regions, resulting in higher uncertainty. These targets are not necessarily less interesting; rather, they are areas where additional drilling would be especially informative for reducing model uncertainty.

This interpretation is consistent with value-of-information and efficacy-of-information approaches to exploration planning, in which the purpose of additional measurements is not only to confirm known targets, but also to reduce uncertainty in geologically and economically relevant parts of the model volume \citep{mern2022intelligent,scheidt2026optimizing}. In this sense, the proposed workflow should be viewed as complementary to expert geological interpretation: it does not provide an automatic drilling plan, but helps identify where the existing data support a target and where additional information would most improve confidence.

\subsection{Comparison with geophysical targets}

The IP survey provides an important independent reference for evaluating the model outputs. The geophysical interpretation shows that not all chargeability anomalies at Kogodai should be treated as direct indicators of economic copper mineralization, because some responses may reflect non-economic conductive or polarizable rocks. Nevertheless, high chargeability combined with low resistivity remains a useful targeting criterion where it is supported by drilling.

Overall, the model partly agrees with the targets proposed from geophysical data alone, but it also rejects or downgrades some of them. It supports follow-up of selected anomalies, including parts of profile 2 and the northern part of profile 3, while it does not predict significant copper mineralization at several other geophysical anomalies. This comparison suggests that the model is useful not only for confirming geophysical targets, but also for ranking them and identifying cases where an IP response may not correspond to a strong copper target.

Where the model highlights additional areas not emphasized in the geophysical interpretation, these should be treated as hypotheses for geological review rather than direct drilling recommendations. Such cases may reflect useful information from historical drilling or spatial trends, but they remain less certain where geophysical coverage or direct sampling is limited.

\subsection{Geological meaning of the predictions}

Geologically, the model was expected to provide two complementary checks. First, it should reproduce the known mineralized zones and their broad geometry. Second, after learning the spatial and geophysical patterns associated with these zones, it should transfer this information to less explored parts of the volume and highlight plausible extensions or new targets.

Within the limits of the available data and model resolution, the workflow identified all recognized mineralized zones within the geophysical data coverage (Zones 1--4), which is an encouraging result because the prediction was data driven rather than manually constrained by zone interpretations. The model also highlighted additional targets recommended for verification, including Zone 2 (yellow) and Zone 5 (peach) in Figure~\ref{fig:mineralization}. Subsequent geological review indicated that these areas are meaningful prospective zones based on field observations and contextual information that were not included in the numerical dataset used for modelling. These results should therefore be interpreted as target-ranking and hypothesis-generation outputs, not as resource-continuity claims.

\subsection{Limitations}

The results should be interpreted in light of several limitations inherent to legacy-driven, multi-source integration. First, historical maps could not be used directly as quantitative model inputs. Many available maps were schematic or lacked sufficient georeferencing information, metadata and spatial consistency for reliable incorporation into the 3D modelling workflow. They were therefore used only as qualitative geological context rather than as evidential layers in the numerical model. As a result, the predictions may underrepresent lithological contacts, structures and other map-derived controls on mineralization that could otherwise help constrain local target geometry.

Second, the geophysical archive contains incomplete metadata and heterogeneous products. In some cases, it was necessary to infer whether files corresponded to pseudo-sections, inversion outputs, horizon-wise slices or derived grids. Although this uncertainty was mitigated by cross-checking against independent constraints and by retaining multiple IP representations where available, incomplete provenance reduces confidence in detailed anomaly geometry. This is one reason why the model relies primarily on relative geophysical contrasts and spatial patterns rather than on absolute values from a single product.

Third, drillhole assays are not continuous geochemical logs. Core sampling was selective and focused on intervals with visible sulfide mineralization, with shoulder samples before and after mineralized zones. This sampling strategy is appropriate for exploration and resource work, but it introduces bias into machine-learning training because sampled intervals are enriched in geologically interesting material. The model therefore learns from mineralization-focused observations rather than from uniformly sampled drillhole trajectories.

Finally, structured geological core descriptions could not be fully exploited as predictor variables. We compiled lithology, alteration, fracture intensity, brecciation, mineralization style and magnetic properties from core logs, but logging standards varied substantially between campaigns and contextual information was incomplete. These attributes were therefore not used in the final modelling stage. This highlights a common bottleneck for machine-learning integration in exploration: model performance is often limited less by algorithmic capacity than by the consistency, completeness and provenance of the underlying geological data.

\subsection{Implications for future work}

Future work should focus on improving the geological and geophysical constraints that most directly affect target confidence. Additional IP/resistivity acquisition should be considered in areas where the model predicts targets but existing geophysical coverage is absent or ambiguous. As recommended in an unpublished geophysical technical report \citep{lvov2025geophysics}, new geophysical data should be delivered in open formats with full metadata, including acquisition geometry, processing history, inversion parameters, coordinate reference information and clear distinction between pseudo-sections, inversion sections, slices and derived grids.

A second priority is systematic data standardization. All drillhole, trench, geophysical, geological-map and assay datasets should be tracked in a consistent spatial database with documented provenance and version control. This would reduce ambiguity in future modelling and make the workflow more reproducible.

A third priority is improved and validated core logging. More consistent lithological, alteration and structural descriptors could substantially improve model quality, particularly because lithology and structure are major controls on mineralization at Kogodai. A focused structural reinterpretation, supported by expert review of fractured and brecciated intervals, may help clarify the role of faults and shear zones as potential ore-fluid pathways. Such work would strengthen the geological basis of future predictive models and improve the reliability of drill-target ranking.

\section{Conclusions}
\label{sec:conclusions}

We developed and tested a 3D drill-targeting workflow for the Kogodai Cu prospect that integrates IP/resistivity data, drilling and trench assays within a Bayesian deep-learning framework. The approach produces spatial predictions of Cu together with epistemic uncertainty estimates, enabling uncertainty-aware prioritization of targets. We consider this an innovative formulation of the exploration problem: rather than forcing sparse drilling, trench and geophysical observations into a simple co-located regression problem (by value interpolation), the workflow learns mutually constrained, spatially continuous 3D fields from incomplete and unevenly distributed evidence. To our knowledge, the use of Bayesian neural networks in this way---to generate continuous 3D predictive fields for deposit-scale drill targeting while simultaneously estimating uncertainty---has not been widely demonstrated in previous mineral-exploration studies.

The credibility of the workflow is supported by its capacity to reconstruct the known mineralized zones without explicit guidance from geological interpretations, zone boundaries or resource-model geometry. In this sense, the model acts as an independent data-driven check on established geological understanding. Its predictive performance is also supported by the threshold-sensitivity and ROC analyses: the Cu prediction separates ore-bearing from ore-free sampled blocks efficiently, with an AUC of 0.93, although this performance should be interpreted most confidently around the more densely sampled parts of the prospect.

The resulting 3D model delineates a principal mineralized trend and several localized candidate zones that coincide with elevated IP responses, while uncertainty mapping highlights where predictions are robust versus where additional data are required. This spatial agreement is geologically meaningful because IP chargeability and resistivity are sensitive to sulfide mineralization, alteration and host-rock contrasts that are relevant to Cu prospectivity at Kogodai. The fact that most predicted targets coincide with IP anomalies supports the interpretation that the model is learning geologically plausible patterns rather than purely statistical artefacts. By applying cutoffs to the continuous Cu-grade prediction, the model can also be expressed as a family of binary prospectivity maps, providing a direct check on how target definition changes with the selected prediction threshold.

The main limitations reflect the legacy nature of the exploration archive: some geophysical data products are not accompanied by full acquisition and processing metadata, and core-logging conventions varied between historical campaigns. Addressing these limitations through standardized open-format data management, improved geophysical metadata and coverage, and harmonized geological logging is expected to substantially strengthen subsequent targeting and interpretation. Despite these limitations, the proposed uncertainty-aware continuous-field approach represents a significant contribution to future drilling planning because it combines target generation, geological validation and risk-aware prioritization within a single reproducible 3D framework.

\printcredits

\section*{Declaration of competing interest}
The authors declare that they have no known competing financial interests or personal relationships that could have appeared to influence the work reported in this paper.

\section*{Data availability}
The drillhole, trench and geophysical data used in this study were provided by Kogodai Joint Venture LLP and are not publicly available due to commercial confidentiality.

\section*{Acknowledgements}

This research did not receive any specific grant from funding agencies in the public, commercial, or not-for-profit sectors.

\bibliographystyle{cas-model2-names}
\bibliography{reference}

\end{document}